\documentclass[aps,prd,
superscriptaddress,nofootinbib,amsfonts,amssymb,amsmath,onecolumn,11pt]{revtex4-1}

\usepackage{color}
\usepackage[usenames,dvipsnames]{xcolor}
\usepackage[normalem]{ulem}
\usepackage{hyperref}
\usepackage{setspace}
\usepackage{orcidlink}
\usepackage{cleveref}
\usepackage{comment}
\usepackage{dsfont}
\usepackage{bm}
\usepackage{enumerate}
\usepackage{placeins}
\usepackage{xurl}

\crefname{section}{Sec.}{Secs.}
\Crefname{section}{Sec.}{Secs.}
\crefname{table}{Tab.}{Tabs.}
\Crefname{figure}{Tab.}{Tabs.}
\crefname{figure}{Fig.}{Figs.}
\Crefname{figure}{Fig.}{Figs.}
\crefname{appsec}{App.}{Apps.}
\Crefname{appsec}{App.}{Apps.}
\crefname{equation}{Eq.}{Eqs.}
\Crefname{equation}{Eq.}{Eqs.}

\hypersetup{
colorlinks=true,
citecolor=blue,
citebordercolor=red,
linktoc=all,
linkcolor=blue,
urlcolor=blue
}

\newcommand{\Observable}{O}
\newcommand{\SampleSize}{N}
\newcommand{\CSetSize}{n}
\newcommand{\CSet}{C}
\newcommand{\Mean}{\mu}
\newcommand{\Std}{\sigma}

\newcommand{\Min}{{\rm min}}
\newcommand{\MeanFitExp}{\alpha}
\newcommand{\StdFitExp}{\beta}
\newcommand{\MeanFitFac}{a}
\newcommand{\StdFitFac}{b}
\newcommand{\Spectrum}{\mathcal{S}}

\newcommand{\NonManifoldlikeness}[1][]{\delta_{\rm #1}}
\newcommand{\ie}{i.e.\,}
\newcommand{\cf}{cf.\,}
\newcommand{\eg}{e.g.\,}
\newcommand{\D}{\mathrm{d}}
\newcommand{\segmentRatio}{segment ratio\,}
\newcommand{\SegmentRatioSymbol}{\rho}
\newcommand{\segmentAngle}{segment angle\,}
\newcommand{\SegmentAngleSymbol}{\theta}
\newcommand{\rotationAngle}{rotation angle\,}
\newcommand{\RotationAngleSymbol}{\gamma}
 \newcommand{\NumLayersSymbol}{N_{\rm L}}   
\newcommand{\AtomsPerLayerSymbol}[1]{n_#1}
\newcommand{\LinkProb}{\Probability_{\rm L}}
\newcommand{\DegreeDist}[2][]{%
  \if\relax\detokenize{#1}\relax
    P_{#2}%
  \else
    P^{\mathrm{#1}}_{#2}%
  \fi
}
\newcommand{\Connectivity}{c}
\newcommand{\Degree}{d}
\newcommand{\Id}{\mathds{1}}

\newcommand{\GraphLap}[1]{\Delta_{\mathrm{#1}}}
\newcommand{\LapEig}{\lambda}
\newcommand{\Multiplicity}[1]{M(#1)}
\newcommand{\Height}{\mathcal{H}}
\newcommand{\SprinkDensity}{\rho}
\newcommand{\Dim}{d}
\newcommand{\HistDummy}{j}
\newcommand{\SumDummy}{i}
\newcommand{\SumDummyTwo}{j}
\newcommand{\ChebyCoeff}{a}

\newcommand{\ChebyDecayBase}{r}
\newcommand{\DegreeMat}{\mathcal{D}}

\newcommand{\ExpectationValue}[1]{\langle #1\rangle}

\newcommand{\RandomVariable}{X}
\newcommand{\NumAtoms}[1]{N_{#1}}

\newcommand{\Num}[1]{n_{\mathrm{#1}}}
\newcommand{\Genus}{g}
\newcommand{\Ord}{O}
\newcommand{\PathLen}{\ell}
\newcommand{\MaxPathLen}{\ell_{\max}}

\newcommand{\Distance}{D}

\newcommand{\Energy}{E}
\newcommand{\Kurtosis}{\gamma_2}
\newcommand{\Skewness}{\gamma_1}
\newcommand{\Probability}{\mathcal{P}}
\newcommand{\CheegerFunc}{\phi}
\newcommand{\CheegerConstant}{\phi}
\newcommand{\DualCheegerFunc}{\bar\CheegerFunc}
\newcommand{\DualCheegerConstant}{\bar\CheegerConstant}
\newcommand{\Mink}{\mathbb{M}}
\newcommand{\Eucl}{\mathbb{E}}
\newcommand{\Median}[1]{\mathop{\mathrm{med}}\limits_{#1}}
\newcommand{\CSetClass}{\mathcal{C}}
\newcommand{\Ensemble}{E}

\newcommand{\Adj}{A}
\newcommand{\LinkMatrix}{L}
\newcommand{\CausalMatrix}{C}
\newcommand{\Graph}{G}

\definecolor{update_color}{HTML}{2E8B57}

\begin{document}

\title{Charting causal set configuration space with graph 
observables}

\author{Astrid Eichhorn\,\orcidlink{0000-0003-4458-1495}}
\email{eichhorn@thphys.uni-heidelberg.de}
\affiliation{Institute for Theoretical Physics, Heidelberg University, Philosophenweg 12 and 16, 69120 Heidelberg, Germany}

\author{Harald Mack\,\orcidlink{0000-0001-7787-9496}}
\affiliation{Scientific Software Center, Interdisciplinary Center for Scientific Computing, Heidelberg University, 69120 Heidelberg, Germany}


\author{Kim Tuyen Le\,\orcidlink{0000-0003-4075-1979}}
\affiliation{Scientific Software Center, Interdisciplinary Center for Scientific Computing, Heidelberg University, 69120 Heidelberg, Germany}

\author{Fabian Wagner\,\orcidlink{0000-0001-8178-9434}}
\email{f.wagner@thphys.uni-heidelberg.de}
\affiliation{Institute for Theoretical Physics, Heidelberg University, Philosophenweg 12 and 16, 69120 Heidelberg, Germany}

\begin{abstract}
The configuration space of causal sets is vast. It is a critical goal to map out this space. Here, we take a practical step towards this goal. We investigate nine classes of causal sets, most of them not studied before. These include manifoldlike causal sets with inhomogeneous Ricci curvature, both topologically trivial and nontrivial. We also study classes of non-manifoldlike causal sets, including lattices, layered orders as well as Lorentzian quasicrystals. Finally, we study classes of causal sets that are not manifoldlike, but are expected to become manifoldlike under a suitable coarse-graining process. We use this broad range of distinct classes of causal sets as a testbed for observables. Rather than focusing on continuum-geometry inspired observables, such as  curvature invariants, which often exhibit large fluctuations and are computationally very expensive, we focus on graph observables, including some observables that constitute subgraph statistics and some that are global. We find that three observables, namely the link degree distribution, the eigenvalues of the graph Laplacian of the symmetrized Hasse diagram and the recently proposed abundance of causal intervals, can distinguish between the distinct classes of causal sets. This is made possible by the small fluctuations that these observables have in most classes.
\end{abstract}

\maketitle

\section{Introduction\label{sec:intro}}
In causal set quantum gravity, the quantum theory of spacetime one aims to construct is the Feynman sum-over-histories of causal sets. A key challenge to overcome is the vastness of the underlying configuration space, i.e., the vast number of possible histories. Causal sets are defined by a set of requirements that -- viewed from the perspective of continuum Lorentzian manifolds -- are the natural requirements for a causal order, combined with spacetime discreteness. Yet, from the perspective of partially ordered sets, the requirements are not particularly strong, in the sense that the resulting space of all causal sets is much larger than that of discretizations of manifolds and contains many classes of physically ``uninteresting" causal sets. These ``uninteresting" causal sets are not discrete counterparts of physically relevant four-dimensional Lorentzian manifolds. More precisely, they have vanishingly low probability of arising from a sprinkling process into a Lorentzian manifold.\footnote{Sprinkling refers to a random distribution of causal set elements into a spacetime according to a Poisson distribution, and a subsequent inference of the relations in the causal sets from the causal order of the continuum spacetime, followed by a removal of all information on the embedding.}
Thus, the set of natural requirements for the causal order of a discrete spacetime is not constraining enough to select the subset of all causal sets that one would ultimately like to retain.

One can take two different attitudes at this point. First, one may reduce the configuration space by constraining the allowed configurations further and selecting only those that are physically interesting. Second, one may not constrain the configurations further, but instead hope that in the sum-over-histories, all physically uninteresting causal sets interfere destructively \cite{Loomis:2017jhn,Mathur:2020hxl,Carlip:2022nsv,Carlip:2023zki,Carlip:2024uny} and therefore do not contribute to observables.\footnote{In a Euclidean setting, a change of measure in a sum-over-histories can in principle be mimicked by a change of action, because $e^{-S}$ acts as a real suppression factor for individual configurations. In a Lorentzian setting, the action removes configurations in sets which interfere destructively, because $e^{iS}$ is never a real suppression factor for real $S$.}

In both attitudes, one must, however, be able to distinguish those causal sets which are sprinklings from those which are not. In the first attitude, this is necessary in order to ensure that, e.g., numerical simulations of the sum-over-histories \cite{Surya:2011du,Glaser:2017sbe,Cunningham:2019rob} stay within the restricted configuration space. In the second attitude, this is necessary in order to understand how to construct a probability amplitude for each configuration that results in destructive interference of the non-sprinklings.

This broader background motivates our specific study. We set out to define and numerically simulate various types of causal sets and subsequently test various observables to determine which are powerful enough to distinguish sprinklings and non-sprinklings.

Observables in causal sets can be constructed following two distinct strategies. The first strategy takes inspiration from continuum differential geometry and gravity. Then, one would want, after simpler global observables like spacetime dimensionality \cite{Myrheim_1978,Meyer_1988,Reid:2002sj,Eichhorn:2013ova,Eichhorn:2019uct}, most importantly find discrete counterparts to the seventeen Zakhary-McIntosh curvature invariants \cite{Zakhary:1997xas} that can be constructed from the Riemann tensor without additional derivatives. These characterize the geometry of four-dimensional spacetime manifolds in classical gravity. Of these, only the simplest one, namely the Ricci scalar, has been constructed \cite{Benincasa:2010ac}; additionally, $\Box\, R$ can be constructed \cite{deBrito:2023axj}. It has been conjectured in \cite{deBrito:2023axj}, that additional higher-order invariants may be encoded in so-called stacked order intervals, but no explicit construction of, e.g., the Kretschmann scalar, has so far been achieved. We call the class of such continuum-differential-geometry-inspired observables \emph{geometric observables}.\\
In the second strategy, one abandons inspiration from the continuum, and focuses more directly on the structure of a causal set as the directed acyclic graph that can be fully encoded in an adjacency matrix or a link matrix. For such graphs, label-invariant subgraph statistics constitute one example of a class of observables that is routinely used in many research areas in which graphs make an appearance \cite{Milo_2002,Kriege_2020}, but which may not have a continuum limit that corresponds to a physically interesting observable on a differentiable manifold. We call the class of such observables \emph{graph observables}.

In this paper, we mainly focus on the second strategy, in the hope of finding graph observables which are a) effective at distinguishing different classes of causal sets and b) calculable within reasonable time in actual numerical studies of causal sets. In the best of all cases, such graph observables may even correspond to geometric observables, because each geometric observable must have a representation as a label-invariant quantity calculable purely from the adjacency matrix (or link matrix). Thus, one might expect that among those graph observables which are particularly powerful at distinguishing distinct classes of causal sets, some have overlap with geometric observables. 

This paper is structured as follows: In \cref{sec:cset_classes}, we introduce the causet classes we investigate. In \cref{sec:GraphObservables}, we present the graph observables. In \cref{sec:results}, we provide our results on the distinguishability of the different classes from causets sprinkled into topologically trivial spacetimes. In \cref{sec:conclusion}, we summarize our findings and conclude the discussion. 

Beyond the main text, in \cref{app:LinkDistMan}, we examine the link-degree distribution of sprinklings into two-dimensional topologically trivial spacetimes more closely. In \cref{app:bound_size_dim} we discuss the size, boundary and dimensionality dependence of the observables. In \cref{app:convergence_plots}, we discuss convergence criteria for our datasets.

The software used for causet creation and computation of observables is available in the repositories QuantumGrav \cite{QuantumGrav}, CausalSetZoology \cite{CausalSetZoology} and CausalSets.jl \cite{CausalSets_jl}. 

\section{Different classes of causal sets}\label{sec:cset_classes}
A causal set $C$ is a set of points $e_i \in C$ with an order relation, $\prec$, called ``precedes", which induces a partial order on the elements of $C$. The physical meaning of the relation $e_i\prec e_j$ is that the spacetime point $e_i$ precedes $e_j$ in a causal sense. A causal set is defined by three requirements, namely
\begin{itemize}
\item[a)] {\bf Transitivity:} If $e_i \prec e_j$ and $e_j \prec e_k$, then $e_i \prec e_k$, $\forall e_i,\, e_j,\, e_k \in C$.
\item[b)] {\bf No closed timelike curves:} If $e_i \prec e_j$, then $e_j \nprec e_i$, $\forall e_i,\, e_j \in C$.\footnote{We use $\prec$ such that $e_i\nprec e_i$, whereas $e_i\prec e_i$ is common in parts of the causal-set literature.}
\item[c)] {\bf Local finiteness:} The causal interval between any two elements of a causal set is finite, $|{e_j: e_i\prec e_j \prec e_k}|<\infty$, $\forall e_i,\, e_k \in C$.
\end{itemize}
The first two conditions are shared between the causal order of spacetime points in continuum manifolds and causal sets, the third condition results in spacetime discreteness.

As highlighted in Sec.~\ref{sec:intro}, while these three conditions are clearly the key conditions that a physically reasonable causal order should satisfy, they are not particularly restrictive. Thus, there are many causal sets which do not correspond to a discrete counterpart of a differentiable manifold at all.

Causal sets can be encoded in the transitive closure or the transitive reduction of directed, acyclic graphs, in which causal set elements correspond to nodes and causal relations to connections between nodes. One can work with the Hasse diagram, \ie, the transitive reduction, which is a graph in which only the links are included, i.e., causal relations $e_i \prec e_j$ for which $|\{e_k|e_i\prec e_k\prec e_j \}|=0$. Links can be denoted as $e_i \prec\! \ast e_j$.
Alternatively one can work with the graph which encodes the causal relations. Note that a generic directed acyclic graph is not a causal set, because it contains only some of the relations implied by transitivity, but not all of them. The transitive closure of any directed, acyclic graph corresponds to a causal set.

Causal sets can also be encoded in matrices. The adjacency matrix is defined by
\begin{equation}
C_{ij} = \begin{cases}
1,\, \mbox{ if } e_i \prec e_j,\\
0,\, \mbox{ else }.
\end{cases}
\end{equation}
Similarly, the link matrix is defined as
\begin{equation}
L_{ij} = \begin{cases}
1,\, \mbox{ if } e_i \prec\!\ast e_j,\\
0,\, \mbox{ else }.
\end{cases}
\end{equation}
Within a natural labeling of a causal set, in which the labeling is compatible with the causal relations, both matrices contain only zeros on and below the diagonal.

A causal set contains various subsets. Particularly important ones are those that correspond to Alexandrov intervals. An Alexandrov interval is a causal interval. The corresponding discrete causal interval can be defined with different conventions regarding whether or not the initial and final point are included. We define the discrete causal interval between $e_i$ and $e_j$ to be
\begin{equation}
I[e_i, e_j] = \{e_k |e_i\preceq e_k \preceq e_j \},
\end{equation}
where $\preceq$ is less strict than $\prec$ in that if $e_i=e_j$, $e_i\preceq e_i$, but $e_i\nprec e_i$. Otherwise, $\preceq$ and $\prec$ have equal meaning. As a result, the cardinality of the smallest causal interval, namely of two elements sharing a link, is 2.

A complete understanding of the configuration space of causal sets does not yet exist. It is known that for causal sets with more than $\CSetSize=80$ elements \cite{Henson:2015fha}, the entropically dominant causal sets are Kleitman-Rothschild orders \cite{Kleitman_1975}, followed by four-layer orders \cite{Dhar_1980,PromelStegerTaraz_2001}. It is also known that an action principle that corresponds to a discrete counterpart of the Einstein action, the Benincasa-Dowker action \cite{Benincasa:2010ac}, can successfully suppress some of these physically uninteresting orders in the sum-over-histories \cite{Loomis:2017jhn,Mathur:2020hxl,Carlip:2022nsv,Carlip:2023zki,Carlip:2024uny}. Yet, many important questions remain open, and we take a step towards addressing some of them. To do so, we define \emph{classes of causal sets} in the following. By a \emph{class}, we refer to a set of causal sets for which a construction principle can be provided, such that a) the construction principle can be implemented in practice, e.g., in numerical simulations, b) the construction principle is specific enough that it results in causal sets which are expected to be similar to each other with respect to a large set of observables and c) the construction principles for different classes are significantly distinct, so that the resulting causal sets are significantly different from each other.
We rush to add that similarity of causal sets can only be properly defined once a (label-invariant) metric on the space of all causal sets is known. As of yet, no such metric exists. In the second part of this paper, when we define observables and investigate how strongly their expectation values differ between different classes of causal sets,  our work may be interpreted as the attempt to construct such a metric, in the spirit of \cite{Bombelli:2000wu,Surya:2025mvt}.

We now define the causal-set classes used throughout this work. We distinguish manifoldlike causal sets generated by Poisson sprinklings into continuum spacetimes from several non-manifoldlike constructions that serve as controlled counterexamples. 
For each class, we specify the generation algorithm and the parameters that vary within the ensemble, so that the observables in \cref{sec:GraphObservables} can be interpreted and compared across classes.

\subsection{Manifoldlike causal sets\label{sec:manifoldlike}}

To define a manifoldlike causal set, we must first clarify how spacetime symmetries shall be treated in the correspondence between a continuum manifold and a discrete causal set. Most importantly, we have to account for the Poincaré group, the symmetry group of Minkowski spacetime. The choice of a discretization necessarily breaks Poincaré symmetry, but one has the option to preserve a discrete subgroup. For instance, by choosing a regular lattice as a discretization, the rotation subgroup and translation subgroup of the Poincaré group are broken to discrete groups, e.g., $\mathbb{Z}_4$ for rotations in a plane. However, there is a highly problematic trade-off between preserving discrete subgroups of the rotation group and breaking Lorentz symmetry by the choice of a preferred frame. Lorentz violations are very tightly constrained by observations \cite{HESS:2011aa,MAGIC:2017vah,MAGIC:2020egb,LHAASO:2024lub}, and therefore the choice of a preferred frame by the discretization is undesirable. To preserve Lorentz invariance in a statistical sense, the correspondence between a continuum manifold and a causal set must therefore be based on a random discretization.

More specifically, a causal set is manifoldlike, if it has a high probability of having arisen from a Poisson process, in which causal set elements are selected from a continuum manifold following the Poisson distribution and the causal relations between causal-set elements correspond to the causal relation inferred from the continuum manifold. 
The Poisson distribution for $n$ points reads
\begin{equation}
    P_V(n)=\frac{(\SprinkDensity V)^n}{n!}e^{-\SprinkDensity V},\label{eq:PoissonSprinkling}
\end{equation}
where $\rho = \frac{1}{\ell^d}$ is the density which defines a discreteness scale $\ell$ in a $d$-dimensional spacetime manifold. $V = \int_R d^dx \, \sqrt{-g}$ is the volume of the spacetime region $R$.
The expected number of points is a direct measure of the volume, \ie, $\ExpectationValue{\NumAtoms{\CSet}}=\rho\, V$. As the volume is Lorentz invariant, so is the average number of points. Therefore, this so-called number-to-volume correspondence is instrumental to satisfying Lorentz invariance \cite{Bombelli:1987aa,Bombelli:2006nm}.

We emphasize that for a causal set to be manifoldlike, it is \emph{not} sufficient that it can be faithfully embedded into a Lorentzian manifold, i.e., it is not sufficient to be able to embed all causal-set elements with their relations in agreement with the causal structure of the manifold. In addition, the distribution of causal-set elements has to obey the number-to-volume correspondence, which is fulfilled for Poisson sprinklings.
Thus, regular lattices, or any other form of regular structure, are not manifoldlike, even if they can be faithfully embedded, because they have vanishing probability to arise from a Poisson sprinkling.

We also highlight that for the correspondence with causal sets, continuum manifolds are grouped into equivalence classes, where two continuum manifolds are equivalent, if they only differ on scales smaller than the discreteness scale. Such differences cannot be resolved by causal sets and are therefore not meaningful within causal set quantum gravity.

\subsubsection{Sprinklings into conformally flat manifolds with varying curvature\label{sec:vary_curvature}}

It is challenging to construct a causal set corresponding to a Lorentzian manifold with non-vanishing Riemann tensor, because, in the absence of sufficiently many Killing vectors and Killing tensors, the null geodesic equation is not integrable. Thus, to determine causal relations between sprinkled points, one must proceed numerically. An exception to this is given by conformally flat manifolds, where the causal relations between two points follow from a flat metric. Therefore, we focus on conformally flat manifolds as the only manifoldlike causal sets that we study. 

This leads to an important limitation of our study, because conformal flatness implies that the Weyl tensor vanishes. Therefore, the Riemann tensor is entirely determined by the Ricci tensor and the Ricci scalar. The severity of this limitation becomes clear, e.g., by noting that of the 17 distinct, non-derivative curvature invariants that characterize four-dimensional Lorentzian manifolds \cite{Zakhary:1997xas}, only two are nonzero for conformally flat manifolds.

On conformally flat manifolds, the Riemann tensor is locally determined by a conformal factor, while the causal structure follows from a flat metric. 
In conformal coordinates $(x_0,\dots,x_{\Dim-1})$, the line element can be expressed as
\begin{equation}
    \D s^2=\Omega(x)^2\eta_{\mu\nu}\D x^\mu\D x^\nu.
\end{equation}
Thus, a $\Dim$-dimensional conformally flat geometry is determined by one function $\Omega(x)$ of $\Dim$ variables.

We sample the conformal factors as an  analytic function of all coordinates. Distinct conformal factors are parameterized by coefficients of an expansion in terms of Chebyshev polynomials. Poisson sprinklings are defined on finite spacetime regions, which motivates working with compact coordinate domains. Chebyshev polynomials provide a natural orthogonal basis on $x_0,\dots,x_{\Dim-1}\in[-1,1]$. For analytic functions, the expansion coefficients decay exponentially with polynomial order. Moreover, higher-order modes encode progressively finer spatial structure, in close analogy to Fourier expansions. As a result, the truncation order provides direct control over the amount of variation in the conformal factor between different points in a spacetime. Thus, the discreteness scale in causal sets is naturally associated to a finite truncation order. For example, in two dimensions we expand the conformal factor in terms of Chebyshev polynomials $T_{\SumDummy}(x)$ as
\begin{equation}
    \Omega(x)=\sum_{\SumDummy,\SumDummyTwo}a_{\SumDummy\SumDummyTwo}\ChebyDecayBase^{-\SumDummy-\SumDummyTwo}T_\SumDummy(x_0)T_\SumDummyTwo(x_1),\label{eq:conformalfactor_Chebyshev}
\end{equation}
where $a_{\SumDummy,\SumDummyTwo}$ are the Chebyshev coefficients, and $\ChebyDecayBase\, \in \mathbb{R}$ denotes the base of the exponential decay.\footnote{The value of $\ChebyDecayBase$ characterizes the ellipse around the origin in the complex plane within which the sampled function is analytic \cite{Boyd01}.} To suppress, rather than enhance, higher-order Chebyshev polynomials, we choose $r$ to be positive and larger than 2.\footnote{Note that we could also use distinct $\ChebyDecayBase_\mu$ for each spacetime dimension. This would not increase the configuration space covered by our parameters, but shift the weight with which we draw geometries from it towards geometries with different scales of fluctuations in different dimensions.}
The Chebyshev polynomials are defined as
\begin{align}
    T_0(x)=1, && T_1(x)=x, && T_{\SumDummy+1}(x)=2\,x\, T_\SumDummy(x)-T_{\SumDummy-1}(x).
\end{align}
To ensure that the conformal factor $\Omega^2$ is positive, we expand its square root in terms of Chebyshev polynomials, which do not have definite sign.
    
In practice, we sprinkle into finite coordinate boxes in conformally Cartesian coordinates, in two or three dimensions. All causal sets of varying curvature are generated using the Julia package \emph{CausalSets.jl} \cite{CausalSets_jl}. In the main text, we limit ourselves to $d=2$ but our prescription generalizes straightforwardly to higher dimensions. We cover the case $d=3$ in \cref{app:bound_size_dim}.

The choice of boundary may affect the expectation value of observables. This is due to the inherent nonlocality of causal relations: for instance, in Minkowski spacetime, the hypersurface spanned by points at spacetime distance 1 to a given reference point is a hyperboloid, i.e., non-compact. Accordingly, the number of nearest neighbors of the reference point in the causal set, i.e., points for which the causal interval with the reference point is 1, is infinite in an infinite sprinkling. Therefore, any choice of boundary removes a subset of the nearest neighbors, which, by Lorentz invariance, are all equivalent and may thus be expected to all contribute to observables. 

Accordingly, our focus on a single choice of boundary is an important limitation of the class of manifoldlike causal sets that we consider.

\subsubsection{Sprinklings into two-dimensional manifolds with varying spatial topology -- generalized pants}\label{sec:pants}
\begin{figure}[!t]
\includegraphics[width=0.5\linewidth]{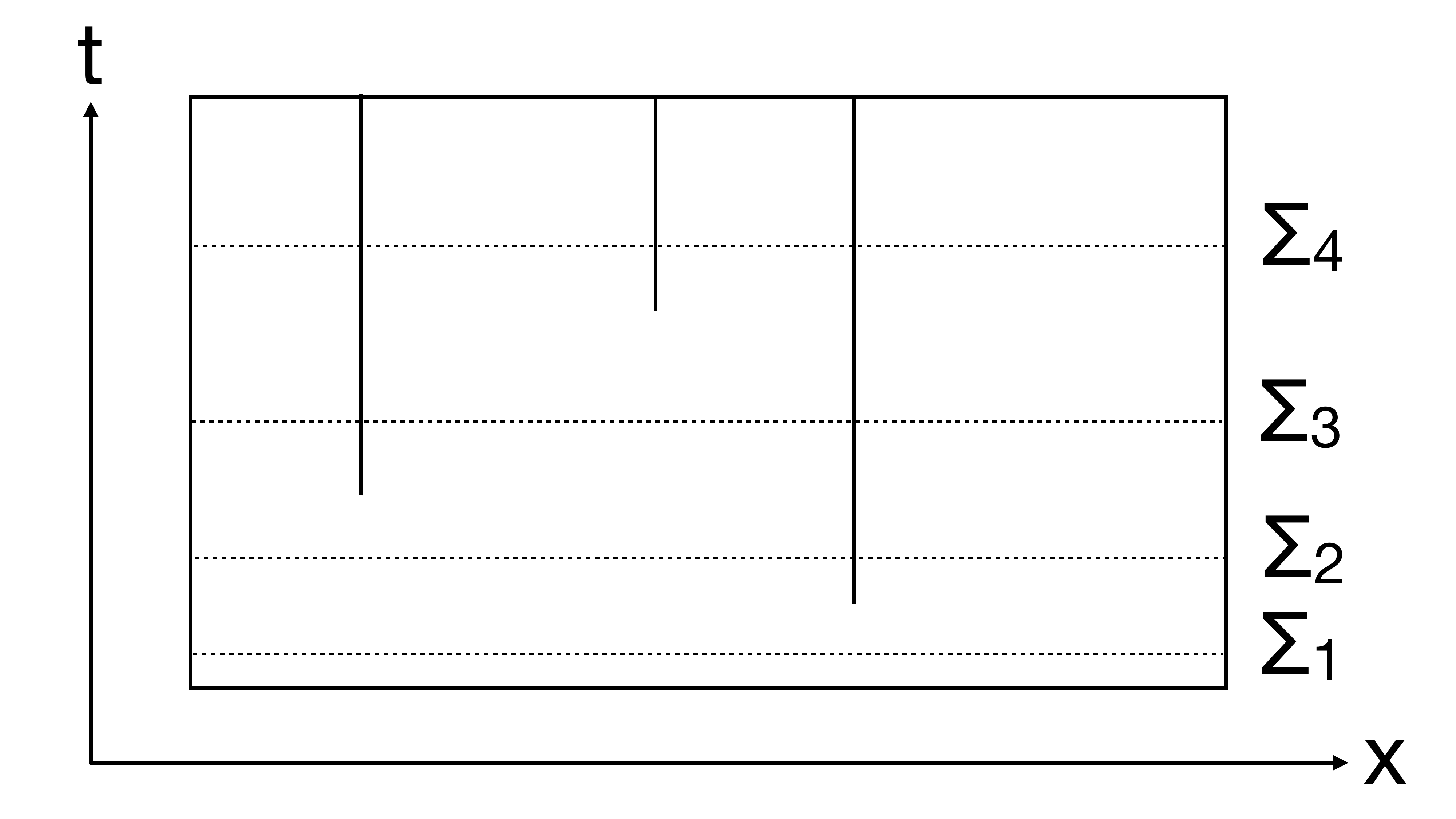}
\caption{\label{fig:pants} We provide an example of a box in 1+1 dimensional spacetime (on which a nontrivial conformal factor can determine the sprinkling density), in which we insert incisions starting from the future spacelike boundary. They result in changes of the spatial topology as a function of time, such that the topology differs between the hypersurfaces $\Sigma_1$ through $\Sigma_4$. Topology in this context refers to whether the spatial hypersurfaces are simply connected or whether they consist of (a varying number of) multiple disconnected pieces.}
\end{figure}

In a box-shaped region in two-dimensional coordinate space, we change the spatial topology on subsequent spacelike hypersurfaces by excising timelike cuts which end on the future spacelike boundary. For a single cut, the region has the well-known 1+1-dimensional pair-of-pants topology. Multiple cuts yield a direct generalization with several branches, see Fig.~\ref{fig:pants}.

Since the cuts form a measure-zero subset of the 1+1-dimensional manifold, the probability for a sprinkled point to be placed on a cut is zero. Thus, the Poisson sprinkling is unchanged. The cuts affect the causal order: we remove causal curves that would intersect a cut, and sever the corresponding causal relations. The net effect is a reduced connectivity between causal-set points in distinct ``legs" of the pants.

In practice, we first sprinkle into a topologically trivial conformally flat geometry of 
varying curvature, i.e., with a spacetime-dependent conformal factor as in \cref{sec:vary_curvature} and then impose the cuts when constructing the adjacency matrix. The position of branch points and therefore the length of the cuts are chosen randomly within the sprinkled boundaries. We only constrain them to be timelike and to connect to the future boundary.

\subsubsection{Sprinklings into two-dimensional manifolds with different Euler characteristic\label{sec:genus}}
Besides changes in spatial topology for manifolds with the spacetime topology of $\mathbb{M}^2$, we can also vary the spacetime topology. In 1+1-dimensions, all possible topologies can be enumerated, and differ in their Euler characteristic $\chi = 2 -2 g$, where $g$ is the genus of the surface.

We thus vary the spacetime topology by adding handles, \ie, by changing the genus $g$ of the underlying two-dimensional manifold, while keeping the boundary of the sprinkled region fixed. 

\begin{figure}[!t]
\includegraphics[width=0.5\linewidth]{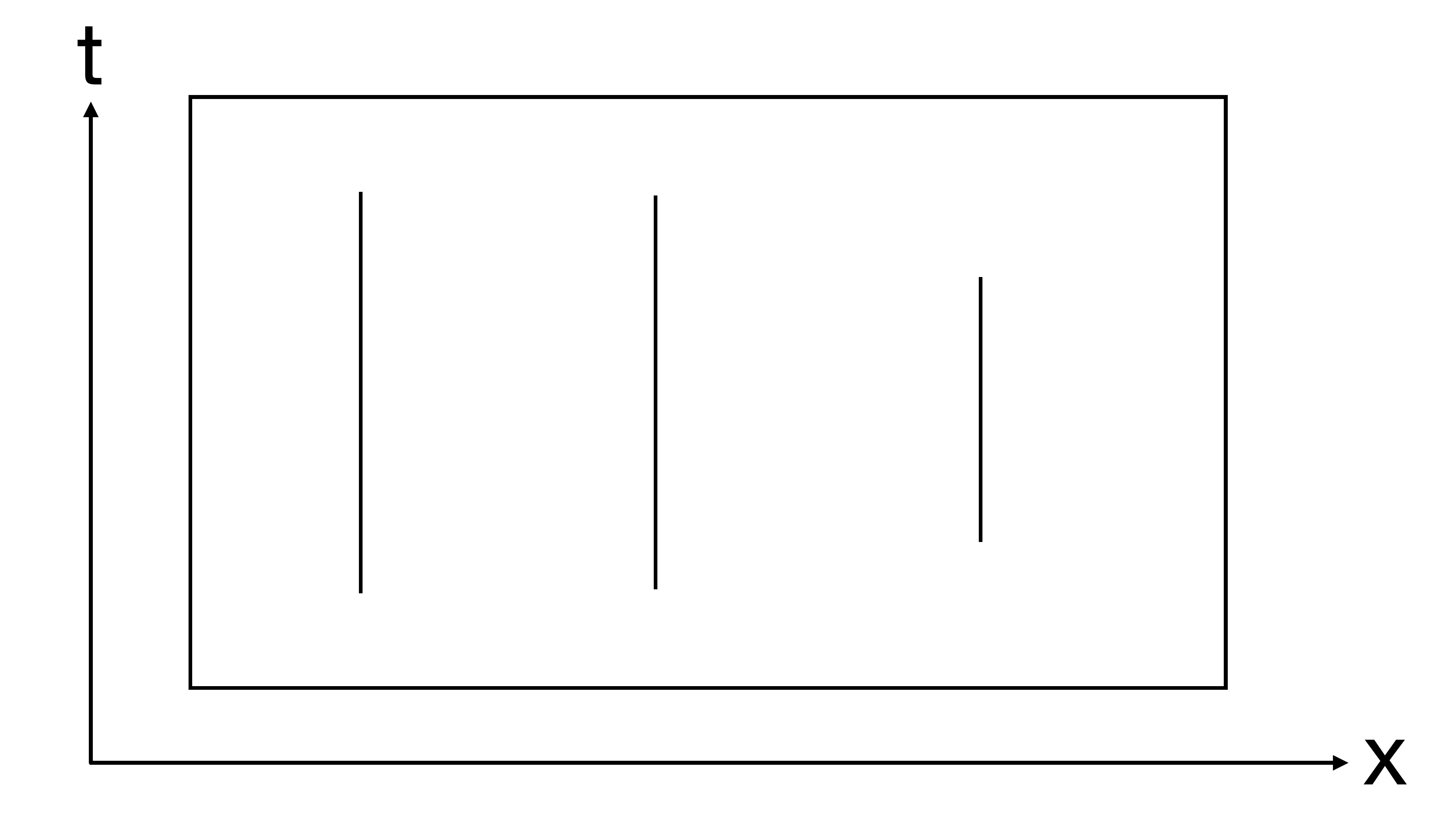}
\caption{\label{fig:Eulercharacteristic} We change the spacetime topology by excising cuts that do not meet the boundary of the spacetime region. This is equivalent to adding handles to a spacetime.}
\end{figure}

We create handles by introducing internal cuts that do not meet the boundary, see Fig.~\ref{fig:Eulercharacteristic}. As before, the excised set has measure zero, so the Poisson sprinkling is unchanged. 
The topology affects the causal order, because the excisions sever relations that exist in the topologically trivial spacetime.

In practice, we sprinkle into a conformally flat geometry with varying curvature. Then, we introduce cuts between pairs of randomly selected points that lie in the open region within the sprinkled boundaries, and compute causal relations taking into account the cuts. The resulting proper length of the cuts, and whether they are spacelike or timelike, is random.

\subsubsection{Summary of classes of manifoldlike causal sets}

Overall, for manifoldlike causal sets we vary the parameters given in Tab.~\ref{tab:manifoldlike_params} independently within the sampling intervals provided there. 

Generally, we work in two spacetime dimensions. In addition, we analyze sprinklings into spacetimes with trivial
topology in three dimensions to discuss the effect of dimensionality.

The causal sets we generate have a fixed size of 2048 elements. For topologically trivial sprinklings, we also compare to samples with differing number of elements in App.~\ref{app:bound_size_dim} to understand the effects of varying the overall size.

Generally, we work with sprinklings into rectangular spacetime regions, i.e., with two spacelike and two timelike boundaries. In \cref{app:bound_size_dim}, we also consider a causal diamond for the topologically trivial spacetimes, in order to understand the effect of changing the boundaries.

We analyze ensembles of $10^4$ causal sets per manifoldlike causal-set type. This number suffices to achieve convergence for the distribution of each observable.
We discuss convergence of our datasets in \cref{app:convergence_plots}.

\begin{table}[]
    \centering
    \begin{tabular}{|c||l l|| c|}
    \hline
        \textbf{symbol}             & \textbf{ meaning}        & &\textbf{sampling interval} \\\hline\hline
         $\ChebyDecayBase$          &  \,inhomogeneity          & Chebyshev-coefficient decay basis. & $[2,8]$  \\\hline
         $\ChebyCoeff_{ij}$         & \,Chebyshev coefficients & characterize conformal factor. & $[0,1]$\\\hline
         $\Num{pants}$              & \,number of pants        & number of cuts ending in boundary. & $[0,10]$ \\\hline
         $g$                        & \,genus                  & number of handles within boundary. & $[0,10]$\\
         \hline
    \end{tabular}
    \caption{Free parameters sampled to generate manifoldlike causal sets with varying curvature and topology. The parameters $r$ and $a_{ij}$ are defined in Eq.~\eqref{eq:conformalfactor_Chebyshev}. We sample flat distributions within each of the sampling intervals. We generate three classes of manifoldlike causal sets. In the topologically trivial class, we vary $(r, a_{ij})$. In the class with varying spatial topology we vary $(r, a_{ij}, n_{\rm pants})$. In the class with nontrivial Euler characteristic, we vary $(r, a_{ij}, g)$.}
    \label{tab:manifoldlike_params}
\end{table}

This results in three distinct classes of manifoldlike causal sets: first, topologically trivial, conformally flat spacetimes with non-vanishing Ricci curvature; second, conformally flat spacetimes with non-vanishing Ricci curvature and varying spatial topology and third, conformally flat spacetimes with nontrivial Euler characteristic.

As both local curvature and nontrivial topology change the number of nearest neighbors of points, we expect that graph observables are sensitive to these changes.

In addition, we expect that the change in connectivity can produce manifoldlike causal sets which are closer (as measured in a given set of graph observables) to some classes of non-manifoldlike causal sets.

\subsection{Non-manifoldlike causal sets\label{sec:non_manifoldlike}}
In this subsection we introduce non-manifoldlike causal-set classes. They violate one or more features expected of faithful sprinklings into smooth, low-dimensional spacetimes, by breaking the number-volume correspondence (which typically results in the selection of a preferred frame), deviating from Poisson statistics, or requiring substructure below the discreteness scale. We study their properties in order to find which structural deviations the observables detect. Besides those non-manifoldlike causal sets which are the most common classes in the configuration space of all causal sets, namely the Kleitman-Rothschild orders and their close ``cousins", the $n$-layer causal sets, we specifically introduce classes of non-manifoldlike causal sets constructed to be close (as measured by a set of observables) to manifoldlike causal sets.

Our overall aim is to map part of the causal-set configuration space, i.e., define classes of causal sets and localize them in the overall configuration space in relation to the other classes. While still far from charting the full configuration space and endowing it with a metric (that would allow to define properly how close distinct configurations are to each other), we take a step in the direction of this goal. 

\subsubsection{Kleitman-Rothschild orders and layered orders\label{sec:KR}}
A Kleitman–Rothschild (KR) order is a finite, partially ordered set that decomposes into three disjoint antichains $L_1$, $L_2$ and $L_3$ \cite{Kleitman_1975}. In the asymptotic limit, $n\rightarrow \infty$, there are $n/2$ elements in $L_2$ and $n/4$ each in $L_1$ and $L_3$.
There are no relations within a layer, and all 
links
are directed 
either from $L_1$ to $L_2$ or from $L_2$ to $L_3$, i.e., there are no links between $L_1$ and $L_3$ and only those relations enforced by transitivity.
Thus, $e_{i}\prec e_j\prec e_k$ $\Rightarrow$ $e_i \in L_1,\, e_j \in L_2$ and $e_k \in L_3$. The probability for a link between any pair of elements in adjacent layers tends to 1/2.

Kleitman–Rothschild orders are non-manifoldlike for several reasons. First, they contain only three ``moments in time". Second, embedding a KR order into a manifold while respecting the causal relations of the order would require highly nontrivial topology, with many of the links corresponding to ``wormholes". In other words, the required manifold would only have an extent of $3 \ell$ in time (in units of the discreteness scale $\ell$) and would have nontrivial topological structure \emph{at that scale}. This is clearly not a structure that we would conventionally refer to as a spacetime manifold.

KR orders dominate the set of finite posets \cite{Kleitman_1975,Henson:2016piq}, \ie, the causal-set configuration space, in the large-$\CSetSize$ limit super-exponentially. Thus, they dominate the sum-over-histories \emph{entropically}. However, one can suppress their contribution \emph{dynamically}:
A nontrivial amplitude, based on the Benincasa-Dowker action \cite{Benincasa:2010ac}, suppresses KR orders in the sum-over-histories \cite{Loomis:2017jhn,Mathur:2020hxl,Carlip:2022nsv,Carlip:2023zki,Carlip:2024uny}. 

KR orders, while dominating the causal set configuration space, are sufficiently different from manifoldlike causal sets, that one may expect that a few simple observables distinguish them. For instance, the fourth power of the adjacency matrix, which counts the number of chains of length four between any pair of elements in the causal set, vanishes for KR orders.
However, they constitute only the proverbial ``tip of the iceberg" of a large set of causets which are all non-manifoldlike, but less straightforward to identify than KR orders.
These are the layered orders.
Layered orders generalize KR orders in that they contain either just two or more than three layers. The case with just two layers is again simple to distinguish. However, for an increasing number of layers, simple diagnostics such as the computation of powers of the adjacency matrix, first becomes computationally expensive, and ultimately fails, because in any finite sprinkling into a manifold, the length of the longest chain is also finite. Therefore, layered orders constitute a class of causal sets for which it is important to find out which computationally viable observable actually allows a distinction from manifoldlike causal sets.

A layered order decomposes into $\NumLayersSymbol\geq 2$ disjoint layers $L_1,\dots,L_k$, each of which is an antichain. Thus, there are no relations within a layer. All relations are directed from lower to higher layers, $L_i \prec L_j$ for $i<j$. Links are allowed only between adjacent layers $L_i\to L_{i+1}$, with transitivity generating relations between non-adjacent layers. 

In our construction, we leave the number of points in each layer $\AtomsPerLayerSymbol{i}$ and the link probability $\LinkProb(i)$ between layers largely unconstrained beyond requiring at least one element per layer. All varied parameters as well as the intervals they are sampled from are summarized in Tab.~\ref{tab:layered_params}.

Mathematically, layered orders are a distinguished class because the number of layers is finite in the limit $\CSetSize\to\infty$. This means that $\NumLayersSymbol/\CSetSize\to0$. Therefore, it is only possible to classify a causal set with finite $\CSetSize$ as layered, if the number of layers $\NumLayersSymbol$ is significantly smaller than the size of the causal set $\CSetSize$. 

Layered orders are non-manifoldlike for the same reasons as the special case of KR orders.

Their relative number in the causal set configuration space decreases with $N_L$, but, at least for low enough $N_L$,  causal sets with $N_L$ layers are the dominant contribution to the causal-set configuration space, once orders with less than $N_L$ layers have been removed.

\begin{table}
    \centering
    \begin{tabular}{|c|| l || c|}
    \hline
        \bf symbol & \bf meaning & \bf sampling interval\\
        \hline \hline
        $\NumLayersSymbol$   & number of layers & $[2,25]$ \\\hline
        $\AtomsPerLayerSymbol{i}$  & number of points per layer $i$ & $[1,\CSetSize-\NumLayersSymbol+1]$\\\hline
        $\LinkProb(i)$ & link probability for pairs of elements in layers $i$ and $i+1$ & $[0,1]$\\
        \hline
    \end{tabular}
    \caption{Free parameters in layered-causal-set generation.}
    \label{tab:layered_params}
\end{table}

\subsubsection{Regular lattice causets\label{sec:grids}}

We construct different kinds of regular lattice causets by embedding the regular grids into a fixed two-dimensional coordinate chart $(t,x)$. We induce a partial order by declaring $i\prec j$ whenever the corresponding points are timelike related with respect to the Minkowski metric on the chart. We obtain Lorentz-violating causal sets, since the point sets single out a preferred frame. They do not satisfy the number-to-volume correspondence, because in the limit $\CSetSize\to\infty$ their regularity leaves infinite volumes devoid of causal-set elements by construction.

\begin{figure}
    \centering
    \includegraphics[width=\linewidth]{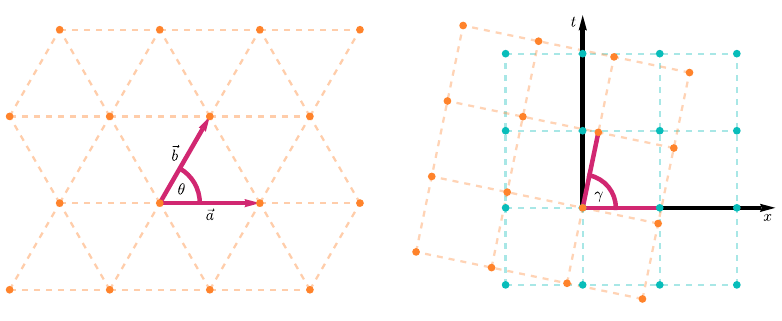}
    \caption{Left: illustration of two characteristic segments of a grid with relative length $\SegmentRatioSymbol=\lVert \vec b\rVert/\lVert \vec a\rVert$ and the enclosed angle $\theta$. Right: Overall rotation angle with respect to $x$-axis $\RotationAngleSymbol$. }
    \label{fig:grid_param_illustration}
\end{figure}

In our construction, we generate a three-parameter family of two-dimensional Bravais lattices, i.e., lattices built by periodically repeating a fundamental cell, inside a fixed coordinate chart $(t,x)$. The parameters and their sampling ranges are summarized in \cref{tab:grid_params}, and illustrated in \cref{fig:grid_param_illustration}. A Bravais lattice is the set of all integer linear combinations of two linearly independent basis vectors. We start from a fundamental cell spanned by two characteristic edge vectors $\vec a$ and $\vec b$. We build the lattices in Euclidean geometry, but then infer the causal relations by interpreting the lattice points as points in Minkowski spacetime. The lattices are parametrized by: \begin{enumerate}[(i)]
\item the ratio of their (Euclidean) lengths, $\SegmentRatioSymbol=\lVert \vec b\rVert/\lVert \vec a\rVert$, 
\item the angle between them $\SegmentAngleSymbol=\angle(\vec a,\vec b)$,  
\item an overall Euclidean rotation $\RotationAngleSymbol$ of the pair ($\vec a,\vec b$) with respect to the $(t,x)$ axes. 
\end{enumerate}
The lattice points are then
\begin{equation}
    (t,x)_{n_1,n_2} = n_1\,\vec a + n_2\,\vec b,\qquad n_1,n_2\in\mathbb{Z}.
\end{equation}
For example, a quadratic grid along the chart $(t,x)$ requires $\SegmentRatioSymbol=1,$ $\SegmentAngleSymbol=\pi/2$ and $\RotationAngleSymbol=0$ such that $\vec a\perp\vec b$ (with either $\vec{a}$ or $\vec{b}$ pointing along the x-axis) and $\lVert a\rVert=\lVert b\rVert$, while a hexagonal grid amounts to $\SegmentRatioSymbol=1$, $\SegmentAngleSymbol=\pi/3$.  To compare to manifoldlike causal sets, we confine the lattices to a rectangular boundary. Finally, we infer causal relations between lattice points from the underlying Minkowski metric. In this step, the two basis vectors can become spacelike, timelike or even null, depending on the choice of angles $\theta$ and $\gamma$.

\begin{table}
    \centering
    \begin{tabular}{|c|| l l || c|}
    \hline
        \bf symbol & \bf meaning & & \bf sampling interval \\
        \hline\hline
        $\SegmentRatioSymbol$  & \segmentRatio: & ratio of the two characteristic edges,  & $[0.1,10]$\\
                               &                &a 2-dimensional Bravais lattice can have. \\\hline
        $\SegmentAngleSymbol$  & \segmentAngle: & angle between characteristic edges of & $[0,\pi/3]$\\
                               &                & two-dimensional Bravais lattice.   \\\hline
        $\RotationAngleSymbol$ & \rotationAngle:& overall rotation of the grid relative  & $[0,\pi]$\\
                               &                & to the coordinate $x$-axis. \\
                               \hline
    \end{tabular}
    \caption{Free parameters in two-dimensional grid generation.}
    \label{tab:grid_params}
\end{table}

Regular lattices can be embedded into Minkowski (or Euclidean) spacetime of suitable dimensionality, and they constitute a simple example of a causal set. Nevertheless, they are not manifoldlike causal sets, because they have vanishing probability of arising from a Poisson sprinkling. As a consequence of the Poisson sprinkling, manifoldlike causal sets, unlike regular lattices, do not select a preferred frame. A key difference between regular grids and sprinklings into Minkowski spacetime is therefore the difference between a size-independent (and typically small) number of nearest neighbors for the lattice, compared to a number of nearest neighbors that increases with the number of points in the sprinkling (due to the non-compact nature of the Lorentz group). 

Regular lattices form an important class within the configuration space, because they  naively appear to be embeddable into Lorentzian spacetimes, but are still non-manifoldlike, as explained above. 

\subsubsection{Spacetime quasicrystals\label{sec:quasicrystals}}

While regular lattices are easy to rule out as the fundamental discrete structure of spacetime, based on Lorentz invariance, spacetime quasicrystals are not. Therefore, we use the recently achieved first construction of spacetime quasicrystals in \cite{Boyle:2026vgo} as our next class of non-manifoldlike causal sets.

Quasicrystals are point sets that are ordered, but not periodic \cite{Penrose1974, Shechtman_1984, Levine_1984, Grunbaum1987, Baake2002}. Unlike regular grids, they lack global translational symmetry and cannot be generated by repeating a fundamental unit cell. Unlike random distributions, they exhibit long-range order because local patterns recur throughout the structure, without repeating periodically. They can be constructed by projecting from a higher-dimensional periodic lattice onto a lower-dimensional physical space \cite{Duneau_1985,Elser_1986,BaakeGrimm2013}. From the projected points, only a discrete subset is selected to form the quasicrystal; we spell out the condition below.
For the resulting pattern to be non-periodic, it is crucial that the lower-dimensional physical space is embedded into the higher-dimensional space at an irrational angle relative to the axes of the higher-dimensional periodic lattice.

\begin{figure}
    \centering
    \includegraphics[width=\linewidth]{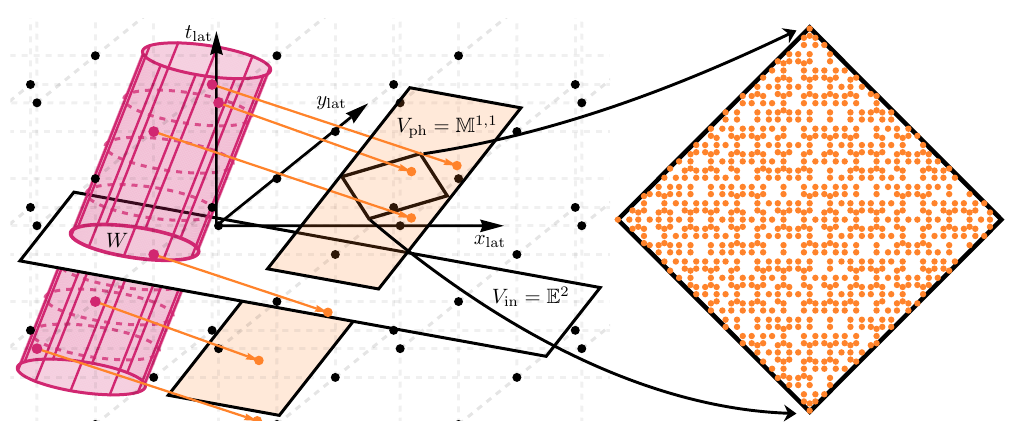}
    \caption{Left: Sketch of projection of hypercubic lattice in Cartesian coordinates in $\Mink^{3,1}$  (one spatial dimension suppressed) to quasicrystal in $\Mink^{1,1}$. Right: Embedding of example quasicrystal in $\Mink^{1,1}$ with causal-diamond boundary.}
    \label{fig:quasicrystal_illustration}
\end{figure}

Spacetime quasicrystals are the analogous construction in Lorentzian geometry, first achieved in
\cite{Boyle:2026vgo}.  We illustrate the procedure in Fig.~\ref{fig:quasicrystal_illustration}. The authors start with a hyper-cubic lattice in Cartesian coordinates in four-dimensional Minkowski space $\Mink^{3,1}$. They partition the space into a Euclidean internal subspace $V_{\rm in}=\Eucl^2$, and a Minkowskian physical space $V_{\rm phys}=\Mink^{1,1}$ such that $\Mink^{3,1}=V_{\rm in}\oplus V_{\rm phys}$. A point on the higher-dimensional square lattice is accepted if its projection onto the internal subspace lands within a disk-shaped acceptance window $W$. If a point is accepted, its projection onto the physical space enters the quasicrystal in $\mathbb{M}^{1,1}$. The radius of the window then determines the density of points of the quasicrystal.

As shown in \cite{Boyle:2026vgo}, Lorentzian quasicrystals are invariant under a dense subset of all Lorentz transformations and preserve the number-volume correspondence. The phenomenological consequences of breaking Lorentz symmetry to a dense subset of all Lorentz transformations are, to the best of our knowledge, unexplored. Therefore, they may at first 
appear to be
a deterministic alternative to Poisson sprinklings. Understanding whether the underlying, higher-dimensional, Lorentz-breaking periodic structure leaves an imprint in observables is therefore crucial.

Interpreted as point sets, spacetime quasicrystals exhibit long-range order:  The Fourier transform of their autocorrelation\footnote{The autocorrelation is the relevant two-point correlation function in diffraction theory, and derived from the point density.} is a discrete set of Dirac peaks \cite{Boyle:2026vgo}. This reflects that correlations of the quasicrystal do not decay at large separations, because patterns reappear at irregularly translated positions. In contrast, the Fourier transform of the autocorrelation of a Poisson sprinkling as a point set is absolutely continuous \cite{Baake_2010,Bjorklund_2022}. It is expected that traces of the long-range order persist in causal sets obtained from quasicrystals. This renders spacetime quasicrystals non-manifoldlike by definition, because a Poisson sprinkling has vanishing probability of producing long-range order.
Still, they can be faithfully embedded into a manifold, respecting both causal structure as well as the number-to-volume correspondence.

In practice, we generate a single large Minkowski quasicrystal restricted to the unit causal diamond, containing $\sim 10^8$ elements. We obtain distinct smaller point sets by selecting causal-diamond subregions of prescribed target size $\CSetSize$, centered at randomly chosen locations within the unit diamond. Causal relations are inherited from the $1+1$-dimensional Minkowski subspace into which we project the elements of the quasicrystal.

\subsubsection{Random partial orders with fixed average connectivity
\label{sec:random}}

Random partial orders with fixed average connectivity mimic the connectivity of sprinklings into topologically trivial spacetimes. The average connectivity $\Connectivity$ is the ratio of the number of relations $\Num{rel}$ and the maximal possible number of relations $\Num{rel,max}=\CSetSize(\CSetSize-1)/2$ for a causet of size $\CSetSize$.

To generate random partial orders with prescribed target connectivity $\Connectivity_{\rm target}$, we use a Metropolis-type Markov chain on the space of finite partial orders. The algorithm starts from an antichain and proposes moves by randomly changing a size-dependent number of entries in the upper triangle of the topologically sorted adjacency matrix, followed by transitive completion. In the first step, this means creating relations, while in later iterations we can both create and remove relations. We accept or reject proposed moves according to a Metropolis criterion with energy
\begin{equation}
    \Energy(\CSet)=|\Connectivity_{\rm target}-\Connectivity(\CSet)|,
\end{equation}
and temperature $T$. Thus, moves that bring the average connectivity closer to the target value are always accepted, while moves that increase the mismatch are accepted with
probability
\begin{equation}   
    \Probability_{\rm acc}=\exp[-(\Energy(C')-\Energy(C))/T] .  
\end{equation}
The procedure stops once the energy drops below $0.1$.

In our datasets, we choose $T\sim 10^{-6}$, which strongly biases the algorithm towards the target connectivity. Thus, the procedure effectively acts as a stochastic optimization algorithm rather than an equilibrium sampler.

\begin{figure}
    \centering
    \begin{minipage}{.49\linewidth}
        \includegraphics[width=\linewidth]{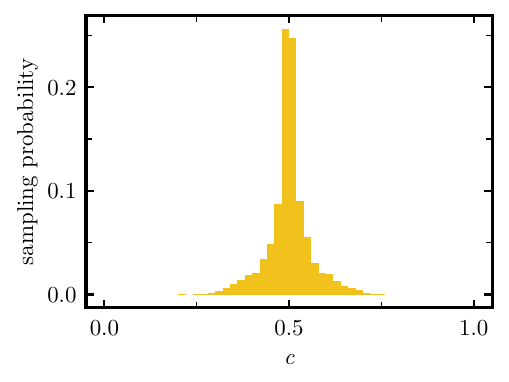}
    \end{minipage}
    \begin{minipage}{.49\linewidth}
        \includegraphics[width=\linewidth]{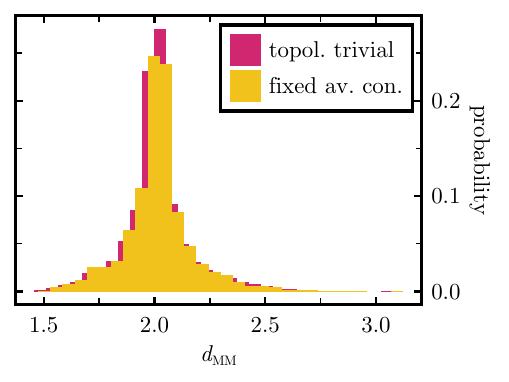}
    \end{minipage}
    \caption{Left: Distribution of average connectivities in two-dimensional topologically trivial spacetimes, empirically obtained from $10^4$ causets of 2048 elements. This is the sampling probability used to generate random partial orders with fixed average connectivity. Right: Distribution over resulting Myrheim-Meyer dimension compared with that of topologically trivial spacetimes obtained from ensembles of $10^4$ causets of 2048 elements.}
    \label{fig:connectivity_man_hist}
\end{figure}

For our datasets, we choose the target connectivities randomly, following the distribution of average connectivities previously obtained for sprinklings into topologically trivial spacetimes, which we provide in the left panel of \cref{fig:connectivity_man_hist}. This ensures that the average connectivity of random partial orders as an ensemble matches that of sprinklings by construction. This also means that the Myrheim-Meyer dimension of the random causets of fixed average connectivity is indistinguishable from that of topologically trivial spacetimes, see right panel of \cref{fig:connectivity_man_hist}. 

It is not obvious that this procedure based on random changes of relations leads to a well-defined causet class. Our results in \cref{sec:results}, however, suggest that the resulting ensemble exhibits limited variability under the observables we consider.  We expect that the class strongly depends on the details of our construction algorithm, i.e., the starting point (antichain), temperature and the numbers of relation changes per iteration, which we fix. In other words, if we, e.g., start from a different causal set and/or significantly alter the number of relation changes per iteration, we expect that the final set of causal sets may differ significantly from ours.

We do not expect random partial orders to be manifoldlike, because the relations are chosen randomly, instead of following the causal order of a spacetime manifold. We construct this class specifically to prevent the average connectivity (or the Myrheim-Meyer dimension) from becoming a simple way of distinguishing manifoldlike causets from non-manifoldlike ones (e.g., regular lattices). As far as we are aware, this class has not been constructed nor characterized before. 

\subsection{Causal sets which may coarse-grain to manifoldlike causal sets\label{sec:almost_manifoldlike}}

In causal-set configuration space, a particularly interesting region is that of causal sets which are non-manifoldlike, but which we expect to coarse-grain to a manifoldlike causal set under a suitable coarse-graining procedure. Coarse-graining procedures have not been systematically defined for causal sets, and, due to the nonlocality that their Lorentzian structure entails, are expected to be significantly different from coarse-graining procedures in, e.g., statistical-physics-models such as the Ising model. Nevertheless, we expect that a consistent coarse-graining-procedure should average over the properties of causal-set elements in a suitable way. Therefore, we expect that an adjacency matrix which corresponds to a manifoldlike causal set, and that is then modified in a small enough subset of its entries, should coarse-grain to an adjacency matrix of a manifoldlike causal set that embeds into the same manifold as the original causal set. 

If such a coarse-graining procedure can be defined, it is then a matter of debate whether the non-manifoldlike causal sets that can be coarse-grained to manifoldlike ones, are ultimately classified as the former or the latter. In the spirit of ``spacetime foam" at the Planck scale, one may well expect non-manifoldlikeness in a causal set close to the Planck scale, such that it only becomes manifoldlike, once a few coarse-graining steps have been completed. This is close in spirit to the idea that suitable observables detect the onset of a stable, manifoldlike regime that emerges under coarse-graining \cite{Major:2009cw}, also in other quantum-gravity approaches \cite{vanderDuin:2025ydn,vanderDuin:2025hmf}. A suitable coarse-graining procedure for causal sets still needs to be developed, see \cite{Eichhorn:2017bwe} for a proposal.

Irrespective of whether or not such a coarse-graining procedure can be defined, these types of causal sets are expected to be ``close" to manifoldlike ones, assuming a suitable metric on the configuration space of all causal sets, i.e., they form a ``transition region" between strictly non-manifoldlike causal sets and strictly manifoldlike causal sets. Thus, they constitute a particularly interesting testing ground for the observables to be defined below.

We monitor this transition by introducing defects into manifoldlike causal sets. We control non-manifoldlikeness with a single parameter $\NonManifoldlikeness$, the relative size of the defect to the full partial order. We study two kinds of defects: Insertions of KR orders, and random addition and removal of relations.

\subsubsection{Manifoldlike causal sets with small KR-order insertions\label{sec:merged}}

We create a manifoldlike causal set with small KR-order insertion with overall size $\CSetSize_{\rm tot}$ as follows:
\begin{enumerate}[(i)]
    \item We generate a manifoldlike causal set of size $\CSetSize_{\rm man}$ with trivial topology in a rectangular boundary as outlined in \cref{sec:vary_curvature}. We vary the parameters $\ChebyDecayBase$ and $\ChebyCoeff_{\SumDummy\SumDummyTwo}$ in \cref{tab:manifoldlike_params}.
    \item We add $\CSetSize_{\rm KR}$ disconnected causal-set elements at a random point in spacetime. We do this by enlarging the adjacency matrix at a random position in the natural labeling by adding $\CSetSize_{\rm KR}$ columns and rows filled with zeros. 
    \item We turn the newly added, wholly disconnected elements into a KR order, which remains disconnected from other elements. We do this by filling the block-diagonal part of the adjacency matrix in the newly created columns and rows with the $\CSetSize_{\rm KR}\times\CSetSize_{\rm KR}$ adjacency matrix of a typical KR order. 
    \item We randomly connect the KR order to different elements of the causal set. We do this by randomly filling the off-block-diagonal elements of the newly created columns and rows with $1$s with link probability $\LinkProb$.  To prevent creating too many easy-to-detect wormhole-like throats, we fix $\LinkProb=0.1$. \label{item:random_connection_KR}
    \item We transitively complete the resulting adjacency matrix to obtain a proper causal-set adjacency matrix.
\end{enumerate}

Here, the non-manifoldlikeness is given by the cardinality of the KR order relative to the cardinality of the resulting causal set: 
\begin{equation}
    \NonManifoldlikeness[KR]=\CSetSize_{\rm KR}/\CSetSize_{\rm tot}.\label{eq:NonManifoldLikenessKR}
\end{equation}
If $\NonManifoldlikeness[KR]$ is small enough, we expect that a suitable coarse-graining procedure should produce a causal set that is approximated by the same manifold as the original causal set before the insertion of the KR order. Because of step \ref{item:random_connection_KR} in the algorithm above, the insertion of the KR order into the causal set is not local, in the sense that the KR order is not contained in a subregion of the originally sprinkled manifold. Its insertion into the adjacency matrix of the sprinkling can rather be understood as the insertion of a collection of ``wormholes" that establish new connections between regions of spacetime that can lie at large timelike, or even at spacelike distance to each other, in the original manifold. In contrast to spacetimes with nontrivial topology, these topological changes increase the connectivity rather than decreasing it. This clearly differentiates causets with defects from spacetimes with nontrivial topology.

\subsubsection{Manifoldlike causal sets with random insertion/removal of relations\label{sec:destroyed}}

We create manifoldlike causal sets with random insertion/removal of relations by adding/removing random relations to/from manifoldlike causets as follows: 
\begin{enumerate}[(i)]
    \item We construct a manifoldlike causal set of trivial topology of size $\CSetSize$ as described in \cref{sec:vary_curvature}.
    \item We randomly choose $\Num{switches}$ entries in the  upper triangular part of the naturally labeled adjacency matrix,  which we switch ($0\to1$ or $1\to 0$).
    \item  We transitively complete the resulting directed acyclic graph.
\end{enumerate}
The transitive completion reverts switches which remove relations that are not links, while enhancing switches which create links by transitively completing a newly created relation. Thus, this construction tends to create more relations than it removes.

 We expect that this process for generating causal sets results in non-manifoldlike causal sets. Our expectation is based on two arguments: first, in the limit of a large number of switches, we expect that the resulting causal set is likely a  KR order, because a random draw of a causal set of fixed size is most likely to be a KR order, once the number of causal set elements is large enough \cite{Henson:2016piq}. Second, even a small number of random switches clearly removes causal relations that are required by the causal structure of the underlying manifold and adds new relations, which are incompatible with it. The new relations in particular relate spacetime points which were spacelike to each other in the sprinkling. It seems clear that even after just a few such moves, adapting the curvature, dimensionality and topology of the underlying manifold above the discreteness scale will not suffice for the causal set to arise with high probability from a sprinkling into such a manifold.

The number of switches $\Num{switches}$ relative to the total number of relations provides a measure of resulting non-manifoldlikeness 
\begin{equation}
    \NonManifoldlikeness[ switches]\equiv\frac{\Num{switches}}{\CSetSize(\CSetSize+1)/2}.\label{eq:NonManifoldLikenessswitches}
\end{equation} 
We expect that a suitably defined coarse-graining reverts the causal set to a manifoldlike one for small $\NonManifoldlikeness[switches]$.

This concludes our discussion of the different causal-set classes we consider. Next, we introduce the graph observables we aim to use to distinguish the classes from each other.

\section{Graph observables and their capability of distinguishing between classes of causal sets\label{sec:GraphObservables}}
Causal sets have long been characterized through observables that constitute discrete counterparts of continuum observables from differential geometry, e.g., the spacetime dimensionality \cite{Myrheim_1978,Meyer_1988,Reid:2002sj,Eichhorn:2013ova,Carlip:2015mra,Abajian:2017qub,Eichhorn:2017djq,Eichhorn:2019uct,Ashmead:2024pmh}, Ricci curvature \cite{Benincasa:2010ac,Benincasa:2010as,Roy:2012uz,deBrito:2023axj}, d'Alembertian \cite{Johnston:2008za,Sorkin:2011pn,Dowker:2013vba,Glaser:2013xha,Aslanbeigi:2014zva,Johnston:2014tia,Belenchia:2015hca,Yazdi:2016eqr,X:2017jal,Yeats:2024tne,Boguna:2025mxz,Kastrati:2025iwv} and others \cite{Major:2006hv,Surya:2007kb,Major:2009cw,Aghili:2018tae,Kambor2020,Machet:2020axq,Mathur:2020hxl,Dowker:2020xfg,Fewster:2020kqo,Adamson:2025fpc}, all reviewed in \cite{Surya:2019ndm, Surya:2025knk}. Yet, taking inspiration from continuum observables may not yield the computationally most efficient as well as quantitatively most precise way to differentiate between different classes of causal sets and distinguish manifoldlike causal sets from non-manifoldlike ones.

Instead, we therefore follow a somewhat different path here, and use observables that are often used to characterize graphs (or networks) in other contexts. These observables, such as, e.g., number of links of a given element of the causal set (also called degree or valency), are of course not unrelated to continuum-inspired observables; e.g., the number of links that a given point has enters the computation of the d'Alembertian, and therefore both the Ricci curvature $R$ \cite{Benincasa:2010ac} as well as the derivative curvature invariant $\Box \, R$ \cite{deBrito:2023axj} at a point. We may even think of the graph observables as candidate building blocks for more complex continuum-inspired observables. 

Because labels on causal-set elements can be thought of as the discrete remnants of coordinates, an observable has to be label-invariant. Similarly, comparing a local observable at a given point in a causal set between different causal sets would correspond to a discrete analogy of diffeomorphism-symmetry breaking. Thus, we compare the distributions of different observables across classes of causal sets, i.e., we use subgraph statistics as well as global statistics to characterize different classes of causal sets.

The observables we choose are complementary to each other:  We choose an observable that captures the local connectivity properties, which is the degree distribution. We choose an observable which captures the distribution of volumes in the causal set, namely the abundance of causal intervals. Further, we choose an observable that encodes global aspects of the connectivity and topology, namely the Laplacian spectrum. Finally, we choose an observable which encodes the global timelike features, which is the height profile. We expect that in combination, these four observables should suffice for a distinction of manifoldlike and non-manifoldlike causal sets within the set of classes that we have defined.
We review these observables, their mathematical definition as well as their physical interpretation below.

\subsection{Link-degree distribution}\label{sec:Degree}
The link-degree distribution is the simplest connectivity observable associated with the links of a causal set. It summarizes how both past- and future-connectivity are distributed across elements. From the point of view of graph theory, this is local information. It is also local with respect to the Lorentzian metric, for which spacetime points at a small distance to each other do not form a compact region.

\subsubsection*{Mathematical definition}

Let $\Graph$ be an $\CSetSize$-element causal set with link matrix $\LinkMatrix$. We define the directed in- and out-degrees as
\begin{align}
\Degree_{\mathrm{out},\HistDummy} \equiv \sum_{\SumDummy=1}^{\CSetSize} \LinkMatrix_{\HistDummy\SumDummy},\quad
\Degree_{\mathrm{in},\HistDummy} &\equiv \sum_{\SumDummy=1}^{\CSetSize} \LinkMatrix_{\SumDummy\HistDummy},
\end{align}
Here, $\Degree_{\mathrm{in},\HistDummy}$ and $\Degree_{\mathrm{out},\HistDummy}$ count the numbers of links into and out of $\HistDummy$.

From the directed degrees we obtain an undirected degree by discarding the orientation of the edges in the graph. For a directed link matrix $\LinkMatrix$ we define
\begin{equation}
\Degree_\HistDummy \equiv \frac{1}{2}\sum_{\SumDummy=1}^{\CSetSize}\bigl(\LinkMatrix_{\HistDummy\SumDummy}+\LinkMatrix_{\SumDummy\HistDummy}\bigr).
\label{eq:DefDeg}
\end{equation}

The degree distribution encodes the probability to find an element with degree $d_j$ in a causal set. Thus, given the set of degrees of all elements in the causal sets, $\{d_1, d_2,...d_n \}$, we define the probability $P_j$ to find $d_i = j$ as
\begin{equation}
\DegreeDist{\HistDummy}\equiv \frac{1}{\CSetSize}\sum_{\SumDummy=1}^{\CSetSize}\delta_{\Degree_\SumDummy,\HistDummy},
\label{eq:DefDegDist}
\end{equation}
where $\delta_{d_i,j}$ is the Kronecker delta. $P_j$ is normalized, 
so that $\sum_{\HistDummy\ge 0}\DegreeDist{\HistDummy}=1$ and therefore is a probability distribution. 

One could also consider the distributions for in- and out-degrees, \eqref{eq:DefDegDist} individually. However, for the causal-set classes considered in this paper, those are highly correlated with the full link-degree distribution, and therefore do not add relevant information.

In principle, the degree distribution can also be defined from the adjacency matrix. The information contained in there is redundant and in practice we find the degree distribution defined from the adjacency matrix to be a relatively noisy observable and do not further study it in what follows.

\subsubsection{Physical interpretation}

For the link matrix, $\Degree_{\mathrm{in}}$ and $\Degree_{\mathrm{out}}$ count immediate predecessors and successors. The link matrix probes local connectivity at the discreteness scale and is sensitive to local geometric as well as topological features and also boundary effects. This is reflected in the link degree distribution.

Link-degree distributions are sensitive to curvature, because curvature changes the continuum volumes of causal neighborhoods, which control the expected number of related (or linked) elements in a sprinkling. Therefore, the number of links at a point also enters the definition of those curvature invariants which have been explicitly constructed for causal sets, namely the Ricci curvature $R$ \cite{Benincasa:2010ac} and its second derivative, $\Box R$ \cite{deBrito:2023axj}. In addition, curvature can affect the in- and out-degree distribution differently. For instance, in de Sitter spacetime, the constant curvature induces systematic shifts between in- and out-degree statistics through the modified volumes of causal neighborhoods.
However, degree statistics are in general not a clean curvature estimator, because inhomogeneous curvature does not lead to a uniform change in the degree distributions.

Similarly, topological effects change the degree distribution, e.g., one can easily imagine how the ``pants" topologies significantly reduce the number of links of a given element within one ``pant leg" compared to a sprinkling in which spatial topology is trivial throughout the original manifold.

For us, it will be relevant to understand whether the changes that spacetime curvature as well as topology can introduce into the degree distribution can or cannot be distinguished from the changes that result from non-manifoldlike behavior.

We expect that the distinction of different classes of causal sets is complicated by boundary effects, because in finite causal sets near a past (future) boundary the out-(in-) degree is depleted.

\subsection{Abundance of causal intervals}
\label{sec:spectrum}
Causal intervals form the backbone of a causal structure. They are precursors to defining aspects of manifoldlike behavior \cite{Glaser:2013pca} and geometric observables \cite{Benincasa:2010ac, deBrito:2023axj}. Nevertheless, the abundance of causal intervals is not an observable that can be defined in the continuum, because one cannot enumerate the set of points contained in a causal interval. In addition, even the relative abundance cannot be defined, because causal intervals can become arbitrarily small in the continuum and thus no overall normalization for the abundance exists. The interval abundance is therefore an example of an observable that falls within what we consider ``graph observables": quantities that characterize graphs, but are not inspired by observables in continuum differential geometry. In \cite{Surya:2025mvt}, the interval abundance has been proposed to not only characterize causal sets, but even define a closeness function on Lorentzian manifolds, using representative causal sets. Previously, causal intervals have also been explored in the context of locality in causal sets \cite{Glaser:2013pca}.

\subsubsection{Mathematical definition}
Given the causal interval $I[e_i,e_j]$ between any pair of elements $e_i \prec e_j$ in a causal set, we can consider those intervals of cardinality $m$, where we remind the reader that we defined the interval $I[e_i,e_j]$ with $e_i \prec\!\ast e_j$ to have cardinality 2. We then define the interval abundance, also called ``spectrum" in \cite{Surya:2025mvt} for a causal set $C$ with adjacency matrix $C_{ij}$ as
\begin{equation}
\mathcal{S}_m = \frac{\CSetSize}{\sum_{i,j} C_{ij}}\sum_{e_i, e_j \in C} \delta_{|I[e_i,e_j]|,m},
\end{equation}
where $\CSetSize$ is the causet size, and $\delta_{|I[e_i,e_j]|,m}$ is the Kronecker delta. Our definition of the interval abundance is normalized such that it provides a probability distribution in the variable $m/\CSetSize\in[0,1]$.\footnote{We choose this normalization because it largely factors out the dependence of the interval abundance on the causet size $\CSetSize$. We discuss the size dependence in \cref{app:bound_size_dim}.} 
\subsubsection{Physical interpretation}
The interval abundance is a simple example of subgraph statistics, because it counts the relative abundance of sub-causal sets of a given size, which are selected due to their causality properties.

Causal intervals are building blocks of the d'Alembertian in a causal set \cite{Sorkin:2007qi,Aslanbeigi:2014zva}, because they can be understood as collecting the ``nearest-neighbors" (links), ``next-to-nearest-neighbors" etc in a causal set. Such neighbor-relations enter the definition of discrete derivatives, and, by extension, the discrete d'Alembertian \cite{Johnston:2008za,Sorkin:2011pn,Dowker:2013vba,Glaser:2013xha,Aslanbeigi:2014zva,Johnston:2014tia,Belenchia:2015hca,Yazdi:2016eqr,X:2017jal,Yeats:2024tne,Boguna:2025mxz,Kastrati:2025iwv} as well as discrete counterparts of commutators of derivatives, i.e., curvature invariants \cite{Benincasa:2010ac,Benincasa:2010as,Roy:2012uz,deBrito:2023axj}.

\subsection{Eigenvalues of the graph Laplacian \label{sec:GraphLaplacian}}

The spectrum of the graph Laplacian provides a compact global summary of connectivity and topology. Eigenvalues diagnose clustering and bottlenecks, while the bulk of the spectrum reflects more homogeneous mixing properties. We use the Laplacian spectrum of the symmetrized link graph, which, being undirected, discards causal orientation.

\subsubsection*{Mathematical definition\label{sec:GraphLapMathDef}}

We construct the symmetrically normalized graph Laplacian for undirected graphs, i.e., graphs with symmetric adjacency matrix $\Adj_S$. It is defined as
\begin{align}
    \GraphLap{\Adj_S}\equiv \Id - \DegreeMat{}^{-\tfrac{1}{2}}\Adj_S\DegreeMat{}^{-1/2},
\end{align}
where  $\DegreeMat_{ij}=\Degree_i\delta_{ij}$ is the degree matrix, which is a diagonal matrix with the degree $d_i$ of the 
element $e_i$ of the graph, defined from the adjacency matrix $\Adj_S$. We take $\DegreeMat^{-\tfrac{1}{2}}$ to be the Moore–Penrose pseudoinverse \cite{Moore_1920,Bjerhammar_1951,Penrose_1955}, so entries of $\DegreeMat$ with $\Degree_i=0$ are mapped to $(\DegreeMat^{-1})_{ii}=0$. This removes contributions from isolated vertices.\footnote{We already capture the abundance of isolated elements in the degree distribution, and therefore do not encode it again in the Laplacian.}

A causal set is not an undirected graph with symmetric adjacency matrix. Thus, to define the graph Laplacian for causal sets, we define a symmetrized link matrix
\begin{equation}
  \LinkMatrix_{\mathrm S} \equiv \LinkMatrix + \LinkMatrix^{T}.
\end{equation}
On the basis of this link matrix, the graph Laplacian can be defined according to
\begin{equation}
\Delta_{L+L^T}= \Id - \DegreeMat{}^{-\tfrac{1}{2}}L_S\DegreeMat{}^{-1/2},
\end{equation}
where the degree is based on the links and defined in \eqref{eq:DefDeg}.

The graph observables that we focus on are the eigenvalues of the graph Laplacian, $\lambda_i$, $i=1,\dots,n$, which are all real \cite{Mohar_1991}. Diagonalizing a dense $\CSetSize\times\CSetSize$ matrix has computational cost $\sim \CSetSize^3$. However, $\LinkMatrix_{\mathrm S}$ is sparse. For sparse, symmetric matrices, one can obtain single eigenvalues at the extreme tails, \ie, the low-lying and high-lying eigenvalues far more efficiently (cost $\sim\CSetSize$) using the implicitly restarted Lanczos method \cite{Calvetti_1994, Sorensen_1997}. Therefore, in \cref{sec:res_lap_eig} we evaluate the capacity of not only the full set of eigenvalues, but also the subset $(\LapEig_2,\LapEig_\CSetSize)$. We choose these values because both are situated at opposite ends of the spectrum. Therefore, we expect that they are complementary, non-degenerate observables, and that they are the most economical to obtain.\footnote{We have verified that $(\LapEig_2,\LapEig_\CSetSize)$ can be obtained for causal sets of size $\sim 10^6$ in 20 seconds on a standard laptop.} Besides, both $\LapEig_2$ and $\LapEig_\CSetSize$ have distinct, and relevant interpretations, which we turn to now.

\subsubsection*{Physical interpretation}\label{sec:GraphLapPhysInt}
The first, critical aspect to point out is that the graph Laplacian does \emph{not} correspond to a physical Laplacian or d'Alembertian on a causal set. This is immediately obvious from the fact that it is insensitive to causal information.

In the context of graphs, diffusion processes can be helpful to probe various properties of the graph, starting from its spectral dimension \cite{Eichhorn:2013ova,Carlip:2015mra,Abajian:2017qub,Eichhorn:2017djq,Eichhorn:2019uct} to the number of disconnected components, as well as properties such as ``small-world-behavior" \cite{Watts_1998,Newman_2000}. This makes eigenvalues of the graph Laplacian powerful graph observables.

We order the eigenvalues as $\LapEig_\SumDummy\leq\LapEig_{\SumDummy+1}$. Without the normalization by $\DegreeMat$, Laplacian eigenvalues are dominated by the degree distribution. The symmetric normalization bounds the spectrum to $\lambda_i\in[0,2]$.

The smallest eigenvalue is $\LapEig_1=0$. 
Its multiplicity $\Multiplicity{\LapEig_1}$ counts the number of disconnected components of the graph. The graph is disconnected if and only if at least two eigenvalues $\lambda_i$ are zero.
Manifoldlike causal sets do not have disconnected components, if the sprinkling is sufficiently dense. Thus, $\Multiplicity{\LapEig_1}\neq 1$ indicates non-manifoldlike structure. 

Eigenvalues which are nonzero, but close to zero, count the number of sparsely connected components of the graph. There are $\HistDummy$ eigenvalues close to zero, if and only if the vertex set contains $\HistDummy$ large subgraphs each of which is only sparsely connected to the respective rest of the graph. This is encoded in Cheeger inequalities \cite{Dodziuk_1984,Alon_1985,Lee_2011}, which bound the eigenvalues from above.

We now make this more precise.
Given a graph $G$, consider $\HistDummy$ pairwise disjoint subgraphs $S_\SumDummy\subseteq\Graph$ with $\SumDummy=1,\dots,\HistDummy$. The set of these subgraphs need not form a partition of $G$, \ie, $S_1\cup\dots\cup S_j\subseteq\Graph$.
For each subgraph $S_i$, we can define the complement $\bar{S}_i = G \setminus S_i$. The number of links between $S_i$ and $\bar{S}_i$ is $e(S_i, \bar{S}_i)$, where the letter $e$ recalls that in standard graph literature, what would be the links in causal sets are typically referred to as ``edges". We define the volume of a subgraph
\begin{equation}
    V(S_i)=\sum_{\SumDummy\in S_i}\Degree_\SumDummy.
\end{equation}
This is not related to the spacetime volume of the region that the corresponding causal set has been sprinkled into.

On this basis, we can
define the expansion of a subgraph $S_i$ as
\begin{equation}
    \CheegerFunc(S_i)=\frac{e(S_i,\bar S_i)}{\mathrm{min}(V(S_i),V(\bar S_i))}.
\end{equation}
$\phi(S_i)$ compares the size of the ``boundary" of $S$ (in the sense of the number of links from $S$ to the rest of the graph) to the smaller value in the pair $(V(S),V(\bar S))$. For a graph in which there are several large clusters,\footnote{ By large we mean that $V(S_i)\sim V(\cup_iS_i)/j$ because then the weight $1/(V(S),V(\bar S))$ is maximal.} which are densely connected \emph{within} the clusters, but not very highly connected \emph{between} the clusters, $\phi(S_i)$ can be close to zero for all $i$, if the subgraphs $S_i$ are chosen to agree with these clusters. For an arbitrary choice of subgraphs $\{S_1,\dots,S_j\}$ of the graph, $\phi(S_i)$ does not necessarily become small for all $i$, because the subgraphs could each contain only a part of the clusters, and the clusters could span several subgraphs. As a result, a set $\{S_1,\dots,S_j\}$ for which the maximal $\CheegerFunc(S_{i_{\rm max}})=\max_{i}\CheegerFunc(S_i)$ is close to zero adequately captures $j$ clusters of $\Graph$.

To properly measure clustering, we need to minimize $\CheegerFunc(S_{i_{\rm max}})$ over all possible sets of subgraphs of $G$, \ie, find the set $\{S_1,\dots,S_j\}$ which best captures the clustering. We define the collection of $j$ non-empty disjoint subsets $\{S_1,\dots, S_j\}$ of $G$, $\mathbb{S}_j = \{ \{S_1,\dots,S_j\},$ $\{S_1',\dots,S_j'\},\dots,$ $\{\{S_1,\dots,S_j\},\emptyset,\dots, \emptyset \} \}$. Each element of the collection, $\mathbb{S}_j$, contains $\HistDummy$ subsets of $G$.
To define $\phi_j(G)$ we find the element $S_{i_{\rm max}}$ with maximal $\phi(S_{i_{\rm max}})$ for each set $\{S_1,\dots,S_j\}\in\mathbb{S}_j$ and then minimize over all $\phi(S_{i_{\rm max}}),$ $\phi(S'_{i'_{\rm max}})$\dots from all sets in $\mathbb{S}_j$. Thus,
\begin{equation}
\CheegerConstant_j=\min_{\{S_1,\dots,S_j\}\in\mathbb{S}_j}\max_{\SumDummy=1,\dots,j}\phi(S_\SumDummy)=\min_{\{S_1,\dots,S_j\}\in\mathbb{S}_j}\CheegerFunc(S_{i_{\rm max}}).
\end{equation}

Higher eigenvalues of the Laplacian satisfy the Cheeger inequalities \cite{Dodziuk_1984,Alon_1985,Lee_2011}
\begin{equation}
    \frac{\LapEig_\HistDummy}{2}\leq\phi_\HistDummy(G)\leq\Ord(\HistDummy^{2})\sqrt{\LapEig_\HistDummy}.\label{eq:Cheeger}
\end{equation}
Thus, a small $\LapEig_\HistDummy$ implies that there is a set of $\HistDummy$ disjoint subgraphs, whose edge boundary to the rest of the graph is small compared to their volume. For $\HistDummy=2$, a small $\LapEig_2$ indicates two large clusters connected by a narrow bottleneck, as expected for a wormhole-like throat.

\begin{figure}
    \centering
    \includegraphics[width=\linewidth]{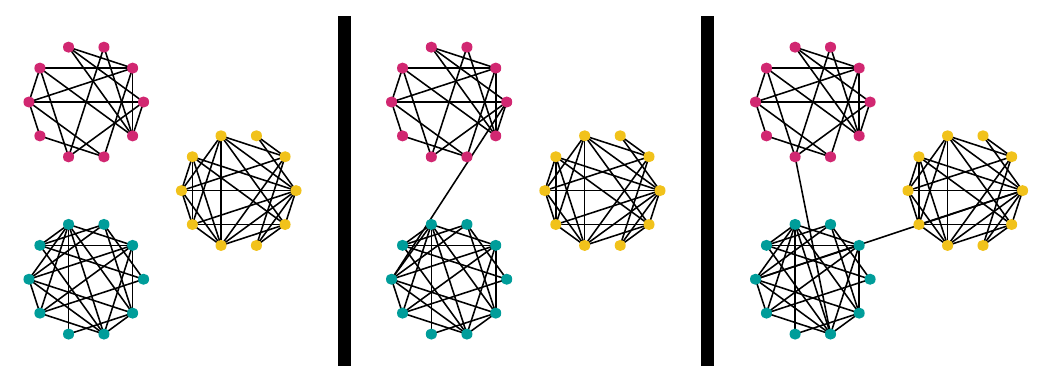}
    \caption{Sketches of three undirected graphs of 30 nodes with three distinguished subgraphs of 10 nodes each. Left: the disconnected components. Middle: two components connected by a one-edge throat. Right: three components connected by one-edge throats.}
    \label{fig:graph_sketch}
\end{figure}

Fig.~\ref{fig:graph_sketch} illustrates the implications of Eq.~\eqref{eq:Cheeger} for three toy graphs. The corresponding multiplicity $\Multiplicity{\LapEig_1}$ and the eigenvalues $\LapEig_2,\LapEig_3$ are listed in Tab.~\ref{tab:graph_sketch}. The multiplicity tracks the number of connected components. Moreover, introducing thin throats suppresses the corresponding eigenvalues, reflecting the presence of weakly connected large subgraphs.

\begin{table}[]
    \centering
    \begin{tabular}{r| c|c|c|}
    \cline{2-4}
         & \textbf{left} & \textbf{middle} & \textbf{right}  \\\cline{2-4}
    $\Multiplicity{\LapEig_1}$ & 3   & 2    & 1    \\
    $\LapEig_2$                & 0.3 & 0.03 & 0.04 \\
    $\LapEig_3$                & 0.6 & 0.3  & 0.05 \\\cline{2-4}
    \end{tabular}
    \caption{Multiplicities of the zero mode and next two eigenvalues for the graphs in Fig.~\ref{fig:graph_sketch}.}
    \label{tab:graph_sketch}
\end{table}

That $\LapEig_\HistDummy$ regards sets of $\HistDummy$ subgraphs implies that $\HistDummy$ is associated with a scale: large $\HistDummy$ reflects properties at small chain lengths, \ie, UV properties, because determining its value requires subdividing the graph into a large number of subgraphs. Small-$\HistDummy$ eigenvalues are global, \ie, IR properties.

The last eigenvalue of the graph Laplacian $\LapEig_\CSetSize$ has its own interpretation. It is bounded by a dual Cheeger-type inequality \cite{Banerjee_2007,Bauer_2009}
\begin{equation}
    2\DualCheegerConstant\leq\LapEig_\CSetSize\leq 2,
\end{equation}
where $\DualCheegerConstant$ is defined as
\begin{align}
    \DualCheegerFunc(S)=&\frac{e(S,\bar S)}{V(S)+V(\bar S)},\\
    \DualCheegerConstant_\HistDummy=&\min_{\{S_1,\dots,S_\HistDummy\}\in\mathbb{S}_j}\max_{\SumDummy=1,\dots,\HistDummy}(\bar\phi(S_\SumDummy)).
\end{align}

When a graph is bipartite, it can be shown that $\DualCheegerConstant=1$ \cite{Banerjee_2007,Bauer_2009}, and thus automatically $\LapEig_\CSetSize=2$. A graph is bipartite if it can be partitioned into two subgraphs which are only related to each other, but have no edges within themselves. In a causal set, this means that there are two perfect antichains whose connections with each other are all links. This is a hallmark of layered structures, as they occur in layered orders (see \cref{sec:KR}).

In short, Laplacian eigenvalues diagnose nontrivial topology in an undirected proxy graph, \ie, they do not encode any causal information. We now complement this acausal global observable with a causal global one that resolves the time ordering of the causal set.

\subsection{Height profile}\label{sec:HeightProfile}
The height profile provides a global summary of the timelike organization of the causal set. It summarizes how elements distribute along a discrete time function defined by maximal chain lengths, and is therefore sensitive to the region’s timelike extent and boundary effects.

\subsubsection*{Mathematical definition}
Let $\CSet$ be a causal set with link matrix $\LinkMatrix$. A directed path from $i$ to $j$ is a sequence
\begin{equation}
    i=e_0\prec\!\ast e_1\prec\!\ast \dots \prec\!\ast e_{\PathLen}=j.
\end{equation}
The path length $\PathLen$ is the number of links.

The maximal path length from $i$ to $j$ is the length of the longest directed path,
\begin{equation}
    \MaxPathLen(i,j)\equiv \max\{\PathLen:\ \exists\ \text{directed path } i\to j \text{ of length }\PathLen\},
\end{equation}
with the convention $\MaxPathLen(i,j)=0$ if there is no directed path from $i$ to $j$.

For each element $i$, we define its largest distance to the future boundary as the maximum over the path lengths from $i$ to any of the elements in the future of $i$,
\begin{equation}
    \ell_{\rm b}(i) \equiv \max_{j\in \CSet}\MaxPathLen(i,j).
\end{equation}

We define the height profile as the normalized histogram of distances to the future boundary of all minimal elements of the causet. A minimal element $e_0$ is an element of a causet $\CSet$ which has no past, \ie, for which $\CausalMatrix_{ie_0}=0$ for all $i$. Thus, the height profile is
\begin{equation}
    \Height_\HistDummy\equiv \frac{1}{\Num{minels}}\sum_{e_0\in\CSet,\CausalMatrix_{ke_0}=0\forall k}\delta_{\ell_{\rm b}(e_0),\HistDummy},
\end{equation}
with the number of minimal elements $\Num{minels}$, and the Kronecker delta $\delta_{\ell_{\rm b},\HistDummy}$,
so that $\sum_{\HistDummy}\Height_\HistDummy=1$.

The maximal height of the causal set is then
\begin{equation}
    \Height_{\max}\equiv \max_{e_0\in\CSet,\CausalMatrix_{ke_0}=0\forall k}\ell_{\rm b}(e_0),
\end{equation}
which equals the maximal cardinality of a chain minus one.

\subsubsection*{Physical interpretation}
The height, in contrast to our other observables, has a clear physical interpretation, namely in terms of the proper time. In causal sets, the proper time between two points is given by the longest chain between them, multiplied by a (dimension-dependent) factor \cite{Brightwell:1990ha}. The usual definition is based on relations, not links, but, because the length of a given chain is longer when any relation within it is substituted by the collection of links from which it follows, discrete timelike geodesics contain only links. Thus, our definition of the maximal distance to the future boundary of a causal set element is the longest proper time a timelike geodesic originating in that element can have within the causet.

This makes $\Height_j$ primarily sensitive to global properties of the sampling region, in particular its timelike extent and boundary geometry. For example, a causal diamond yields a characteristic height profile with few elements at very small and very large heights and a broad maximum in the bulk. This is clearly different for a rectangular boundary.

Similarly, non-manifoldlikeness can leave characteristic imprints, which is most obvious in the case of layered orders, for which the height profile only covers the values $[1,\dots,\NumLayersSymbol -1]$.


\section{Results: Mapping out the unknown: First steps into the causal set configuration jungle\label{sec:results}}
We provide results for the four sets of observables (degree distribution, interval abundance, eigenvalues of graph Laplacian and height profile), compared across the nine classes of causal sets:
\begin{itemize}
\item[I)] {\bf Manifoldlike causal sets:}
\begin{itemize}
\item[a)] Topologically trivial, conformally flat, two-dimensional sprinklings with varying conformal factor, which we denote as {\bf topologically trivial spacetimes}. Within this class, the parameters $r$ and $a_{ij}$ in Tab.~\ref{tab:manifoldlike_params} are varied and we show the resulting expectation value and standard deviation arising from these variations.
\item[b)] Conformally flat, two-dimensional sprinklings with varying spatial topology and varying conformal factor, which we denote as {\bf generalized-pants spacetimes}, parameterized by $r$, $a_{ij}$ and $n_{\rm pants}$ in Tab.~\ref{tab:manifoldlike_params}. Within this class, for each distinct value of $n_{\rm pants}$ that we include in our results,
$r$ and $a_{ij}$ are varied.
\item[c)] Conformally flat, two-dimensional sprinklings with nontrivial spacetime topology, which we denote as {\bf spacetimes with handles}. These are parameterized by $r$, $a_{ij}$ and genus $g$ in Tab.~\ref{tab:manifoldlike_params}.
\end{itemize}
\item[II)]{\bf Non-manifoldlike causal sets that may coarse-grain to manifoldlike ones}
\begin{itemize}
\item[d)] Topologically trivial, conformally flat, two-dimensional sprinklings with varying conformal factor, with the insertion of a KR order, parameterized by $\delta_{\rm KR}$, the  size of the KR order relative to the size of the causal set. We denote these as {\bf spacetimes with KR insertions}.
\item[e)] Topologically trivial, conformally flat, two-dimensional sprinklings with varying conformal factor, in which $n_{\rm switches}$ entries of the adjacency matrix are switched from 0 to 1 and vice versa. They are parameterized by $\NonManifoldlikeness[switches]$, which provides the (square root of the) ratio of switched links to all links. 
Because a resulting ``missing link" is akin to a ``hole" in a causal set and an ``added link" is akin to a wormhole,
we denote these as {\bf spacetimes with (worm-)holes}.
\end{itemize}
\item[III)] {\bf Non-manifoldlike causal sets}
\begin{itemize}
\item[f)] KR orders and layered orders, which are parameterized by the number of layers $N_{\rm L}$, the number of points $n_i$ per layer $i$ and the probability for pairs of elements in layers $i$ and $i+1$ to be linked, $P_{\rm L,i}$. We denote these as {\bf layered causets}.
\item[g)] Regular $k$-valent lattices, which are parameterized by the ratio of the two characteristic distances to the nearest neighbors of a point, $\delta$, by the angle between these, $\theta$, and by the angle $\gamma$ by which the lattice is rotated compared to the $(t,x)$ coordinate axis. We denote these as {\bf regular lattice causets}.
\item[h)] Spacetime quasicrystals, which we denote by {\bf quasicrystal causets}.
\item[i)] Random partial orders with fixed average connectivity, which we denote as {\bf fixed-connectivity causets}. We sample the fixed average connectivity $\Connectivity$ from the distribution in \cref{fig:connectivity_man_hist}.
\end{itemize}
\end{itemize}

In this section, for each observable we show how well the observable distinguishes between the different causet classes. As context for these results, we discuss the dependence of observables on size, boundary and dimensionality for topologically trivial spacetimes in \cref{app:bound_size_dim}, and convergence in \cref{app:convergence_plots}.

We quantify how an observable distinguishes between classes based on a distance function: Large distance indicates good distinguishability. We achieve this as follows: Given an observable $\Observable$ with $\Num{bins}$ bins, we define the distance between two causets $\CSet_1$ and $\CSet_2$ for the observable as
\begin{equation}
    \Distance_\Observable(\CSet_1,\CSet_2)=\sum_{\HistDummy=1}^{\Num{bins}}|\Observable_\HistDummy(\CSet_2)-\Observable_\HistDummy(\CSet_1)|,
\end{equation}
\ie, as the integrated per-bin difference. To determine whether the distance between two causal sets indicates good distinguishability of the classes by the observable, we need to know the variation of $\Distance_\Observable$ within a class. We estimate the variation within a class, or null distance $\Distance_{0,\Observable}(\CSetClass)$, as the median of the distances obtained from pairing all causets in a single-class ensemble $\Ensemble_\CSetClass$
\begin{equation}
    \Distance_{0,\Observable}(\CSetClass)=\Median{\CSet_1\neq \CSet_2 \in \Ensemble_\CSetClass}\Distance_\Observable(\CSet_1,\CSet_2).
\end{equation}
On this basis, we quantify the distinguishability of a causet class $\CSetClass$ from the comparison class $\CSetClass_0$ as the probability to find a pair $(\CSet_1,\CSet_2)$, where $\CSet_1\in\CSetClass$ and $\CSet_2\in\CSetClass_0$, with distance $\Distance_\Observable(\CSet_1,\CSet_2)<\Distance_{0,\Observable}(\CSetClass_0)$, \ie,
\begin{equation}
    \Probability_\Observable(\CSetClass,\CSetClass_0)=\frac{1}{|\Ensemble_\CSetClass||\Ensemble_{\CSetClass_0}|}\sum_{\CSet_1\in\Ensemble_{\CSetClass},\,\CSet_2\in\Ensemble_{\CSetClass_0}}\theta[\Distance_{0,\Observable}(\CSetClass_0)-\Distance_\Observable(\CSet_1,\CSet_2)],\label{eq:distinguishability_probability}
\end{equation}
where $|\Ensemble|$ is the number of causets contained in the ensemble $\Ensemble$ and $\theta$ is the Heaviside theta function. This means that the smallest probability a set of two ensembles $(\Ensemble_\mathcal{C},\Ensemble_{\mathcal{C}_0})$ can achieve is approximately $1/(|\Ensemble_\CSetClass||\Ensemble_{\CSetClass_0}|)$.

If $\Probability_\Observable(\CSetClass,\CSetClass_0)$ is very small, the observable $\Observable$ differs between causal-set classes $\CSetClass$ and $\CSetClass_0$ far more strongly than it varies within $\CSetClass_0$. Thus, given a typical element of $\CSetClass$, we can say with high confidence that it is not in $\CSetClass_0$.

In the following, we generally choose topologically trivial spacetimes as the comparison class.\footnote{Only for quasicrystal causets we use Minkowski sprinklings with causal-diamond boundary as null class because the quasicrystal causets themselves describe Minkowski spacetime with causal-diamond boundary.} This is motivated by the fact that we consider these the physically most relevant and would like to understand which observables are capable of distinguishing this physically more relevant from the physically less relevant classes. 
\subsection{Link-degree distribution}

\begin{figure}[!t]
    \centering
\includegraphics[width=0.9\linewidth]{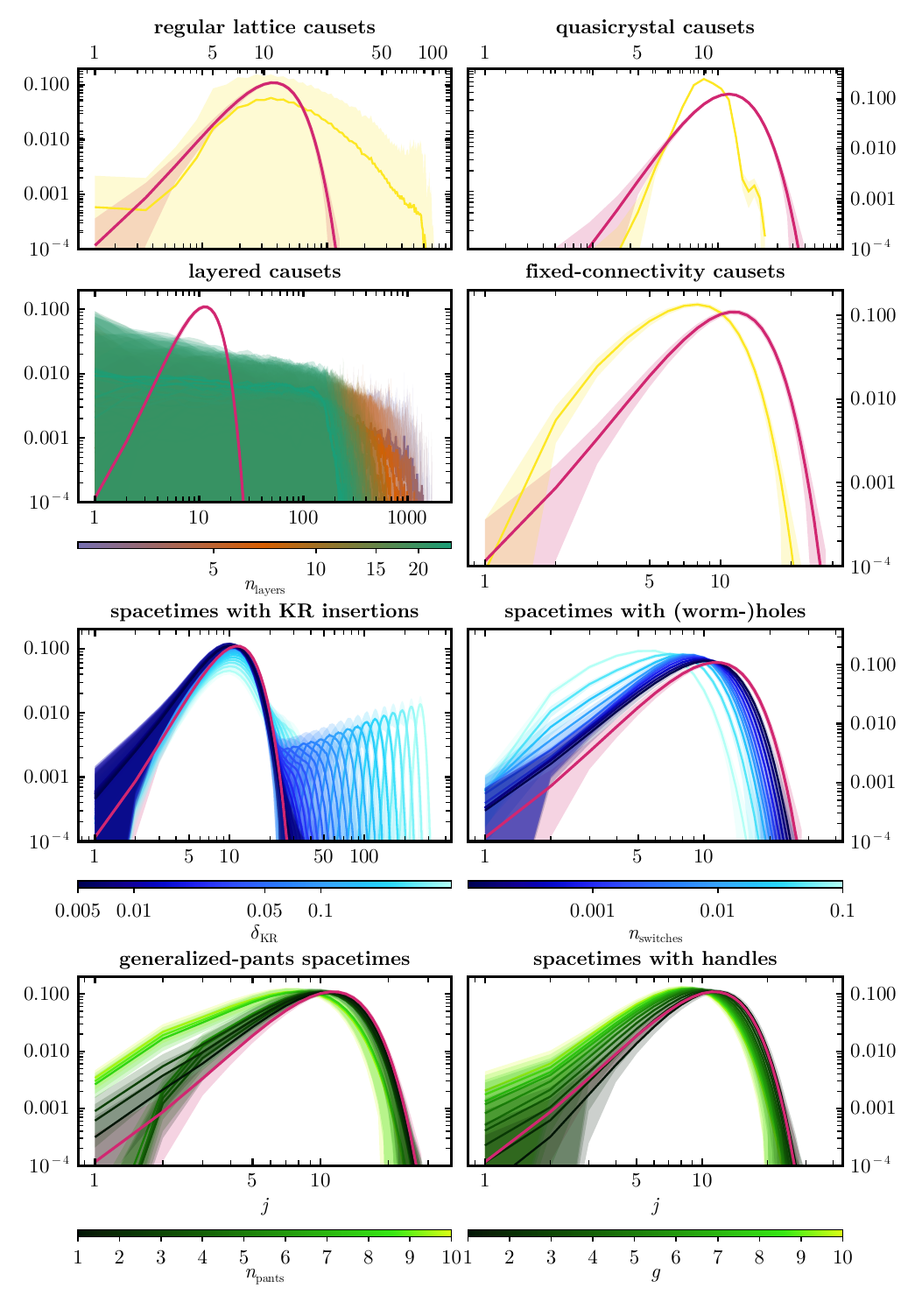}
    \caption{Link-degree distribution for different kinds of causal sets compared to sprinklings into topologically trivial spacetimes (magenta). Lines are means, bands are standard deviations. Datasets contain $10^4$ partial orders for each class ($3\cdot 10^4$ for spacetimes with KR insertions and spacetimes with wormholes) of size 2048.}
    \label{fig:plot_matrix_degrees}
\end{figure}

The results for the link-degree distribution are shown in Fig.~\ref{fig:plot_matrix_degrees}. In each panel, the magenta distribution shows the result for topologically trivial spacetimes. There are several interesting observations to make:

First, the link degree distribution is an observable with little noise, except for the case of layered and regular-lattice causets which, as all observables show, are intrinsically diverse classes. 

Second, the link degree distribution for topologically trivial spacetimes differs significantly from the other classes that we study. This suggests that using a link action in the sum-over-histories, i.e., an action in which the leading term is based on links, may indeed be a first step towards dynamically suppressing non-manifoldlike causal sets through destructive interference, as seen in \cite{Loomis:2017jhn,Mathur:2020hxl,Carlip:2023zki,Carlip:2024uny}.

Third, the link-degree distribution for topologically trivial spacetimes is a Gaussian distribution that is deformed by being skewed towards smaller degrees. The skewness appears to be a pure boundary effect. We make this statement precise in \cref{app:LinkDistMan}, where we discuss the link-degree distribution of topologically trivial causets in more detail. The Gaussian distribution distinguishes topologically trivial spacetimes from other classes, including quasicrystal causets, for which the link degree is \textit{not Gaussian} distributed, but significantly more peaked, see \cref{app:LinkDistMan} for more details.

\begin{table}[!t]
    \centering
    \begin{tabular}{| l| c |}
    \hline\textbf{causet class} & \textbf{probability}\\\hline
    regular lattice causets & $<10^{-8}$\\\hline
    quasicrystal causets & $<10^{-8}$\\\hline
    layered causets & $4\cdot 10^{-4}$ \\\hline
    fixed-connectivity causets & $<10^{-8}$\\\hline
    \end{tabular}
    \caption{Distinguishability probabilities for the link-degree distribution for all causet classes according to \cref{eq:distinguishability_probability} for ensembles of $10^4$ causets of size $\CSetSize=2048$ per class.}
    \label{tab:link_degree_distances}
\end{table}

Fourth, differences to topologically trivial spacetimes become more pronounced as the parameters $\NonManifoldlikeness[KR]$, $\NonManifoldlikeness[switches]$, $n_{\rm pants}$ and $g$, increase. This implies two things: First, the degree distribution is sensitive to such deviations from topologically trivial spacetimes. In other words, it is sensitive to topological properties, as well as non-manifoldlike insertions into manifolds. Second, within each such class, the degree distribution even contains information on the value of the parameter $\NonManifoldlikeness[ KR]$, $\NonManifoldlikeness[switches]$, $n_{\rm pants}$ or $g$, respectively. 

Fifth, for fixed-connectivity causets the degree distribution itself has little noise, suggesting that our algorithm for constructing such causets indeed results in a well-defined class of causal sets. The same conclusion is supported by our other observables below. 

\begin{figure}[!t]
    \centering
\includegraphics[width=\linewidth]{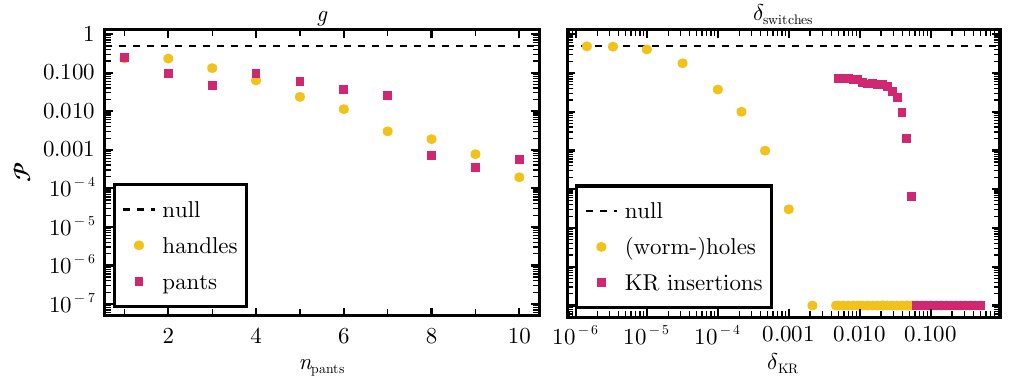}
    \caption{Distinguishability probabilities using the link-degree distribution for spacetimes with handles, generalized-pants spacetimes, spacetimes with (worm-)holes and spacetimes with KR insertions, compared to the expected null value $1/2$ according to \cref{eq:distinguishability_probability} for ensembles of $10^3$ causets of size $\CSetSize=2048$ per point ($10^4$ causets in the null class). The resolution for probabilities is $10^{-7}$, \ie, points with probability $10^{-7}$ are upper bounds rather than exact values.}
    \label{fig:degree_probabilities_panel}
\end{figure}

We quantify how well the link-degree distribution distinguishes all classes from topologically trivial spacetimes according to Eq.~\eqref{eq:distinguishability_probability} in Tab.~\ref{tab:link_degree_distances} and Fig.~\ref{fig:degree_probabilities_panel}. We find that the link-degree distribution can easily distinguish all classes within Tab.~\ref{tab:link_degree_distances} from topologically trivial spacetimes. However, it detects topology and boundary modifications only if those are strong (large $\Genus$ or $\Num{pants}$). Spacetimes with wormholes can be distinguished well, when at least $\sim 0.1\%$ of the links have been modified, while at least $\sim5\%$ of a size $\CSetSize=2048$ causet has to be turned into a KR order for it to be detectable with link degrees.

\subsection{Interval abundances}
\begin{figure}
    \centering
    \includegraphics[width=0.92\linewidth]{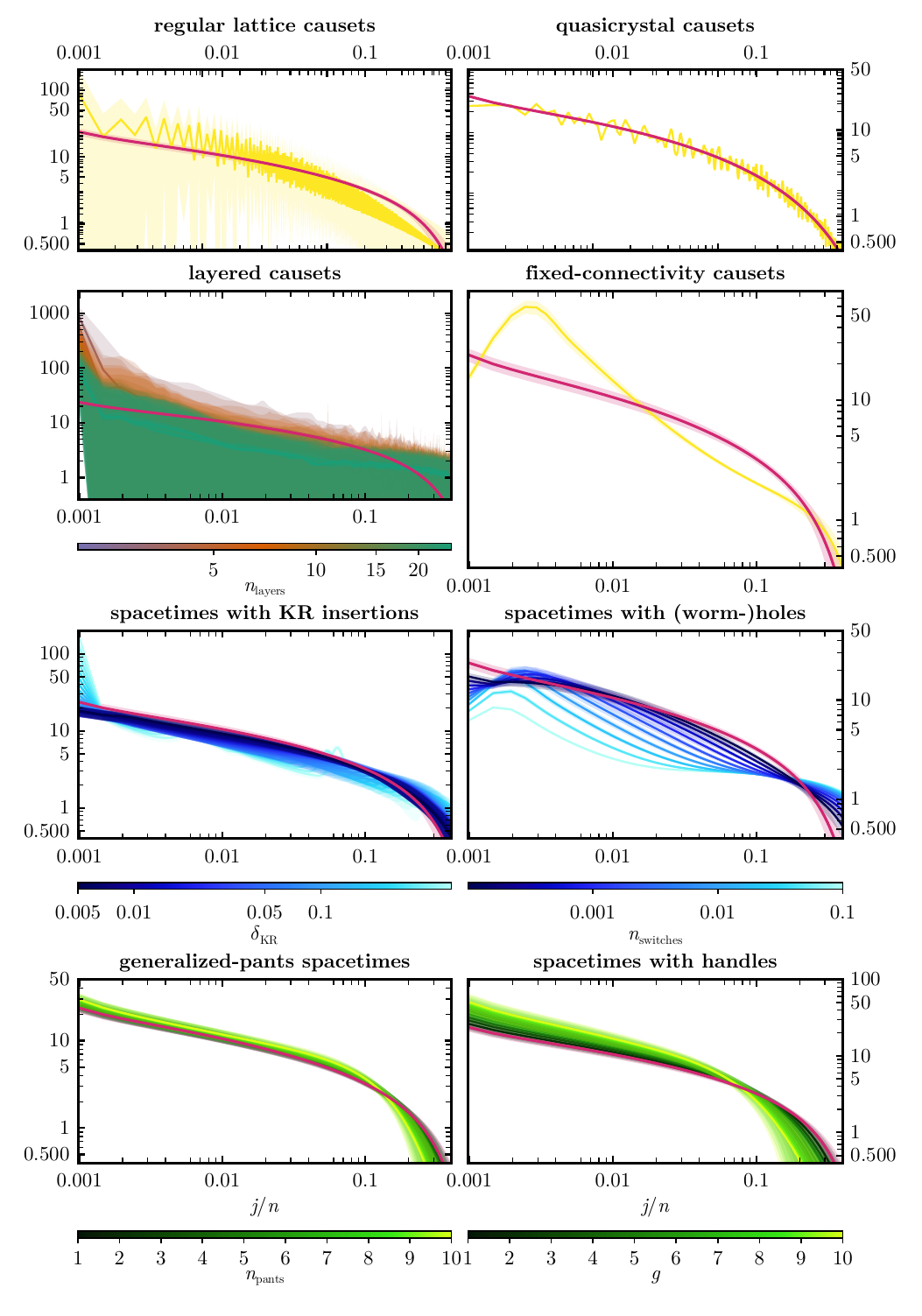}
    \caption{Interval abundances for different kinds of causal sets compared to sprinklings into topologically trivial spacetimes (magenta). Lines are means, bands are standard deviations. Datasets contain $10^4$ partial orders for each class ($3\cdot 10^4$ for spacetimes with KR insertions and spacetimes with wormholes) of size 2048.}
    \label{fig:plot_matrix_cardinalities}
\end{figure}

The interval abundance has been conjectured in \cite{Surya:2025mvt} to be a useful distance function on Lorentzian geometries, even though it is understood to be degenerate on some physically distinct geometries. Here, we extend the studies in \cite{Glaser:2013pca,Surya:2025mvt} to several classes of causal sets not considered in \cite{Glaser:2013pca,Surya:2025mvt}. We show the interval abundances of all causet classes in \cref{fig:plot_matrix_cardinalities}, and observe the following characteristics.

First, the interval abundances have little noise except for regular lattice causets and layered causets.

Second, the interval abundances can clearly distinguish topologically trivial spacetimes from the other types we study. This indicates that derived observables like the Benincasa-Dowker action are also sensitive to class changes as was shown for layered orders in \cite{Carlip:2022nsv,Carlip:2023zki,Carlip:2024uny}.

\begin{figure}[!t]
\begin{center}
\includegraphics[width=\linewidth]{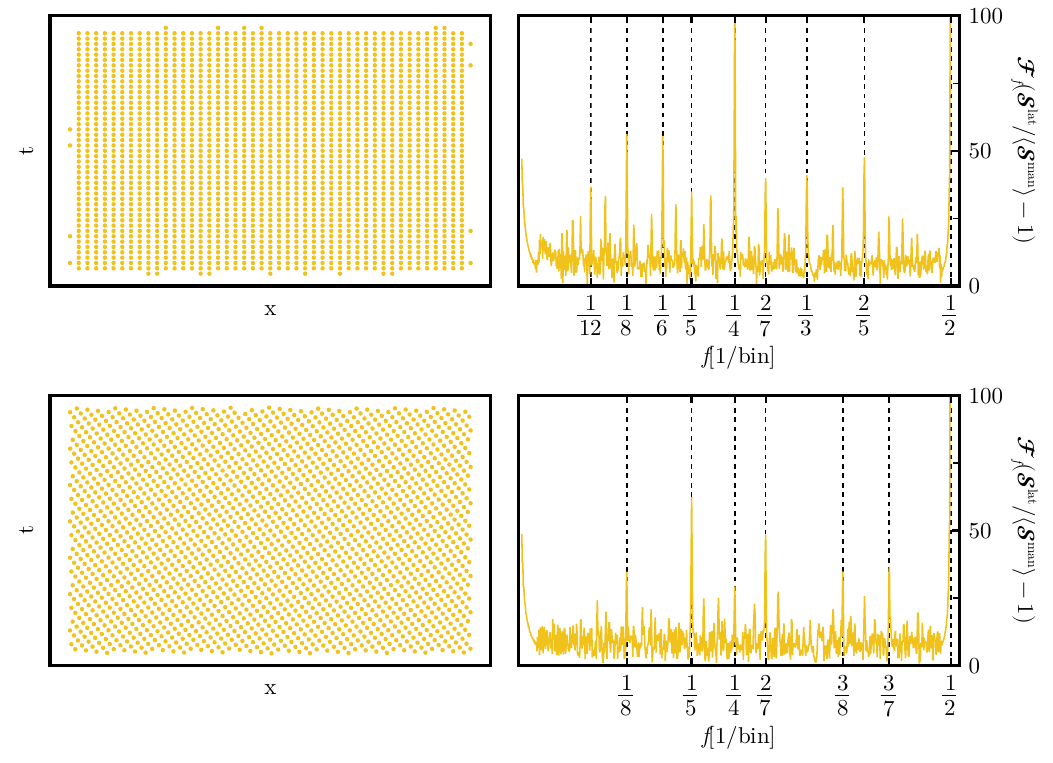}
\end{center}
  \caption{Left panels: We show the regular lattice causets embedded in Cartesian coordinates; right panels: We show the Fourier transform of the interval abundance as defined in \cref{eq:FT}. Characteristic frequencies are highlighted by dashed black vertical lines. All causets have 2048 elements and are constrained to a rectangular boundary. Upper row: Square lattice rotated by an angle $\RotationAngleSymbol\sim10^{-14}$. Bottom row: Oblique lattice with segment ratio $\SegmentRatioSymbol = 2$ and segment angle $\theta=73\pi/180$ rotated by an angle $\RotationAngleSymbol=43\pi/180$.
    \label{fig:FTlattice}}
\end{figure}

Third, for regular lattice causets, we observe a periodicity in the interval abundance.\footnote{The periodicity is easiest to understand for the non-rotated square lattice, where each point has three outgoing links (one timelike one, ``straight up" and two lightlike ones, along the diagonal). Thus, besides the links themselves, there are no causal intervals smaller than 5 elements (we are including the bottom and top element in the count). There are also no causal intervals of size 6 and 7, but intervals of size 8 exist.} This suggests that the (discrete) Fourier transform of the interval abundance is a useful observable to consider in this case. In fact, we first divide the interval abundance by the expectation value for the interval abundance of topologically trivial spacetimes, because we observe that the interval abundance for regular lattice causets appears to follow the same overall trend with oscillations superimposed. Next, we take a discrete Fourier transform to arrive at Fig.~\ref{fig:FTlattice}, which shows
\begin{equation}
\mathcal{F}_f\left(x\right)=\sum_{\HistDummy=1}^{\CSetSize}e^{-i\frac{2\pi(\HistDummy-1)(\CSetSize f-1)}{\CSetSize}}\left(x_j\right)\label{eq:FT}
\end{equation}
for $x=(\Spectrum^{\rm lat}/\ExpectationValue{\Spectrum^{\rm man}})-1$, $x_j=(\Spectrum_j^{\rm lat}/\ExpectationValue{\Spectrum_j^{\rm man}})-1.$
Regular lattice causets admit characteristic frequencies which set them apart from all other causet types we have studied. The characteristic frequencies and associated periods contain information about the lattice parameters in \cref{tab:grid_params}.

\begin{figure}[!t]
    \centering
        \includegraphics[width=.5\linewidth]{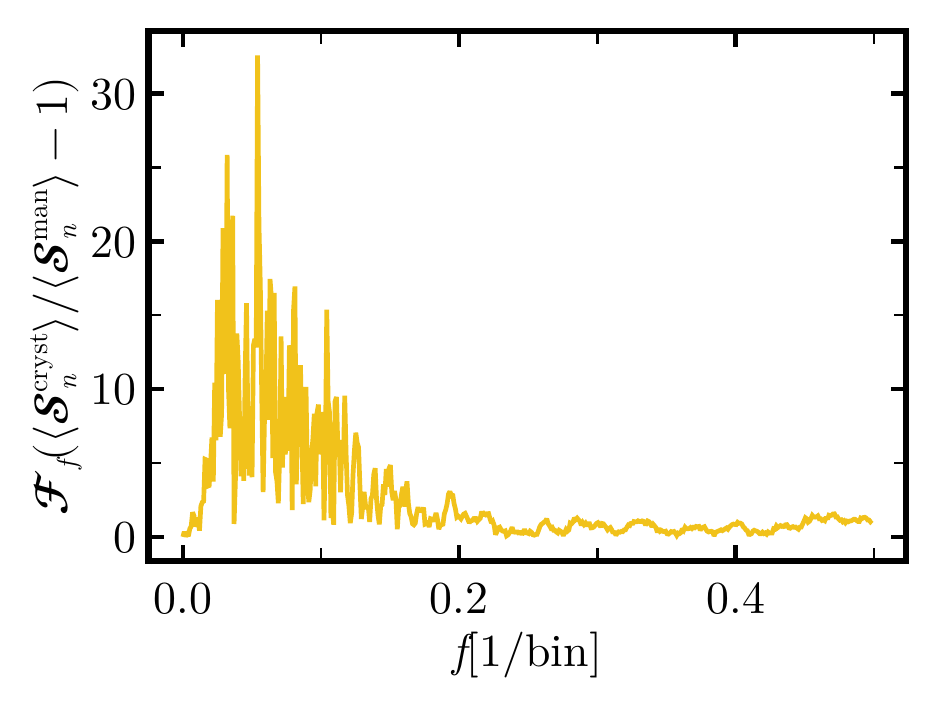}
    \caption{Fourier transform of fluctuation of quasicrystal abundances $\Spectrum^{\rm cryst}_n$ around the abundances of topologically trivial spacetimes $\Spectrum^{\rm man}_n$ as a function of the frequency in inverse numbers of bins for causets of size 2048. Note that in comparison to \cref{fig:FTlattice}, this plot shows the Fourier transform of the ensemble average over $10^4$ distinct quasicrystal causets instead of just a single one.}
    \label{fig:FTquasicrystal}
\end{figure}

Fourth, the interval abundance is sensitive to the (quasi-)periodicity of quasicrystals and is qualitatively similar to that of regular lattice causets, cf.~Fig.~\ref{fig:plot_matrix_cardinalities}.  This motivates us to consider its Fourier transform defined in \cref{eq:FT} for $x=(\langle\Spectrum^{\rm cryst}\rangle/\ExpectationValue{\Spectrum^{\rm man}})-1$, $x_j=(\langle\Spectrum_j^{\rm lat}\rangle/\ExpectationValue{\Spectrum_j^{\rm man}})-1$, shown in \cref{fig:FTquasicrystal}. Here, we consider the Fourier transform over the ensemble \textit{average} interval abundances of quasicrystal causets weighted by the average interval abundances of sprinklings into Minkowski spacetime. We make this choice because quasicrystals are not invariant under global discrete translations like regular lattices. In contrast to lattices, their structural information is not contained in a single, finite causal set. We observe several characteristic frequencies, similar to the case of regular lattice causets. These lie in a smaller interval at low frequencies, compared to the regular lattice causets, which tend to have characteristic frequencies distributed over the entire range $[0,1/2]$.

Fifth, we find that the smallest differences arise for topological modifications of spacetimes, i.e., generalized-pants spacetimes and spacetimes with handles. Even those, however, differ from topologically trivial spacetimes, once $\Num{pants}$ or $\Genus$ is large enough. In turn, the interval abundance is very sensitive to small modifications away from manifoldlikeness, which we expect to coarse-grain to manifoldlike causal sets. For instance, the interval abundance of spacetimes with (worm-)holes differs drastically from that of topologically trivial spacetimes, already at $\NonManifoldlikeness[switches] =10^{-3}$, i.e., once about one in every $1000$ links has been modified. Similarly, an insertion of a small KR order leaves an easily detectable imprint on the abundance of small intervals. This is because a KR order in itself is characterized by a large number of small intervals.

\begin{figure}[!t]
    \centering
\includegraphics[width=0.96\linewidth]{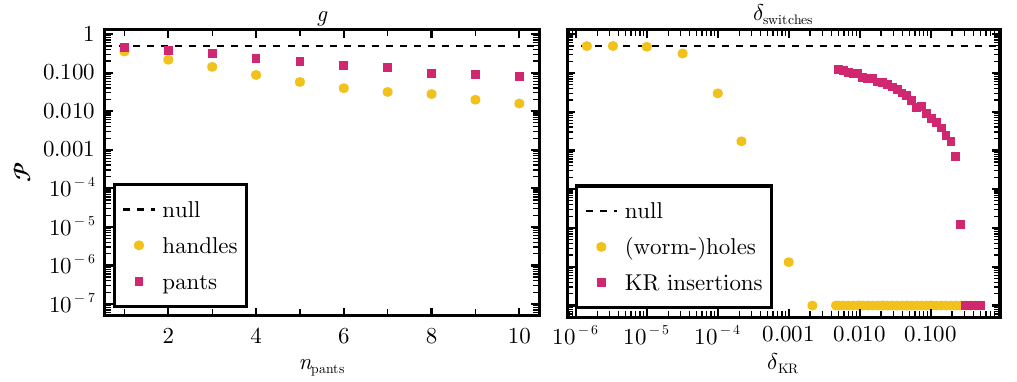}
    \caption{Distinguishability probabilities using the interval abundances for spacetimes with handles, generalized-pants spacetimes, spacetimes with (worm-)holes and spacetimes with KR insertions, compared to the expected null value $1/2$ according to \cref{eq:distinguishability_probability} for ensembles of $10^3$ causets of size $\CSetSize=2048$ per point ($10^4$ causets in the null class). The resolution for probabilities is $10^{-7}$, \ie, points with probability $10^{-7}$ are upper bounds rather than exact values.}
    \label{fig:cardinalities_probabilities_panel}
\end{figure}

In \cref{tab:cardinalities_distances} and \cref{fig:cardinalities_probabilities_panel} we compare between causets in different classes more precisely by asking probabilistic questions about the expected interval abundance for a single causal set in each class. While the interval abundances can clearly distinguish the types in \cref{tab:cardinalities_distances}, they fare comparably badly at identifying topological or boundary modifications. Similarly to the link-degrees, interval abundances detect removed or added links if $\NonManifoldlikeness[switches]\sim10^{-4}$ of the resulting causet have been modified, and require around $\NonManifoldlikeness[KR]\sim5\cdot10^{-2}$ of the causets with KR insertions to consist of KR orders for the effect to be clearly visible.

\begin{table}[!t]
    \centering
    \begin{tabular}{| l| c |}
    \hline\textbf{causet class} & \textbf{probability}\\\hline
    regular lattice causets & $<10^{-8}$\\\hline
    quasicrystal causets & $<10^{-8}$\\\hline
    layered causets & $< 10^{-8}$ \\\hline
    fixed-connectivity causets & $<10^{-8}$\\\hline
    \end{tabular}
    \caption{Distinguishability probabilities for the interval abundances for all causet classes according to \cref{eq:distinguishability_probability} for ensembles of $10^4$ causets of size $\CSetSize=2048$ per class.}
    \label{tab:cardinalities_distances}
\end{table}

\subsection{Graph-Laplacian eigenvalues \label{sec:res_lap_eig}}

\begin{figure}
    \centering
    \includegraphics[width=0.92\linewidth]{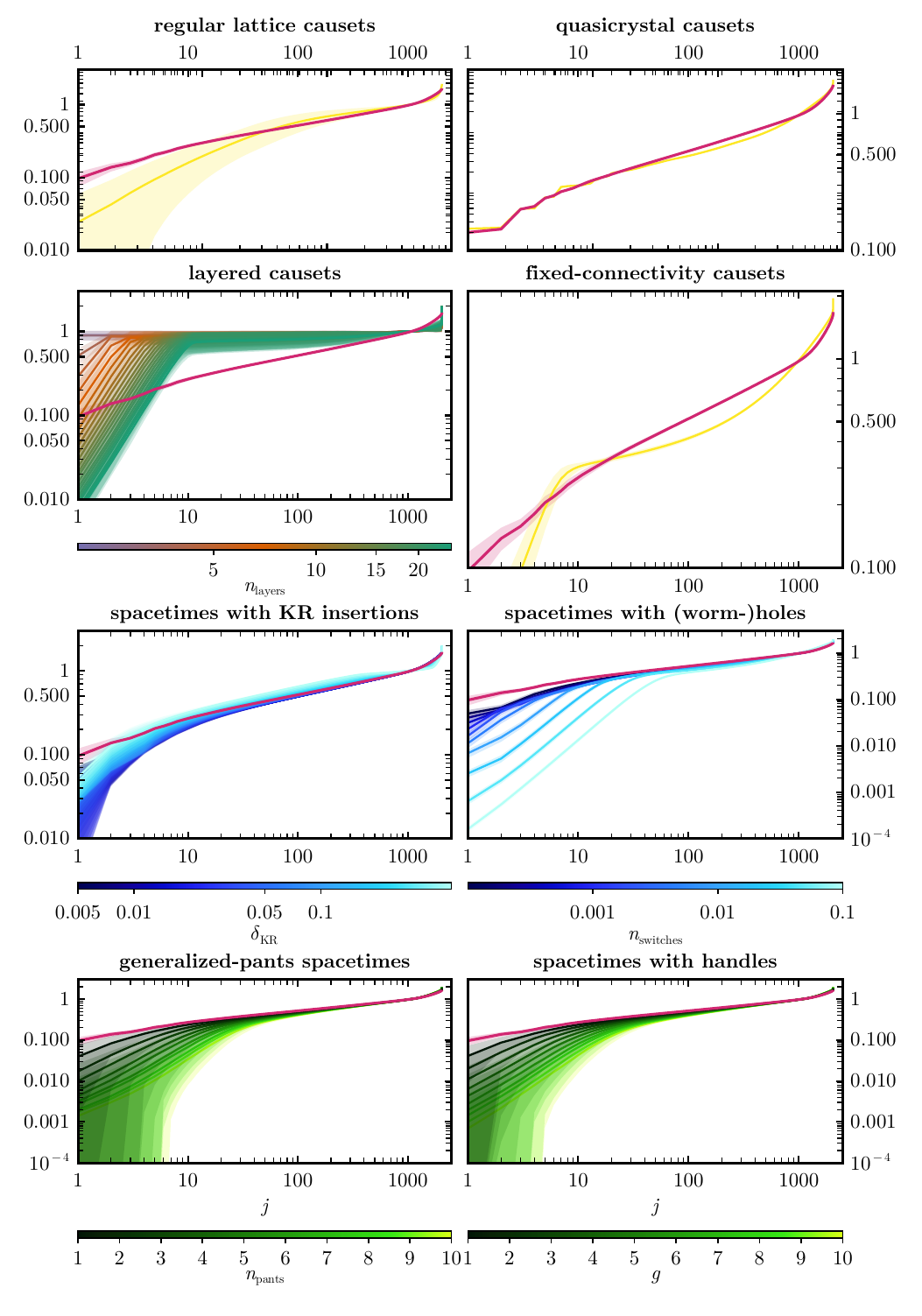}
    \caption{Laplacian eigenvalues for different kinds of causal sets compared to sprinklings into topologically trivial spacetimes (magenta). 
    Lines are means, bands are standard deviations. Datasets contain $10^4$ partial orders for each class ($3\cdot10^4$ for spacetimes with KR insertions and spacetimes with wormholes) of size 2048.}
    \label{fig:plot_matrix_eigvals}
\end{figure}

The eigenvalues of the graph Laplacian are shown in \cref{fig:plot_matrix_eigvals}. We first focus on the low-lying eigenvalues, which contain information on overall connectivity. Manifoldlike causal sets have previously been shown to exhibit ``superdiffusion" \cite{Eichhorn:2013ova}, i.e., a diffusion process on a sprinkling explores the overall causet much faster than it would explore a regular lattice of the same overall size. This follows from the larger overall connectivity that a sprinkling has compared to a regular lattice. In turn, the resulting ``nonlocality" is directly linked to the non-compactness of the Lorentz group and the resulting large spatial distances that links cover in any given frame. Thus, we expect that the lowest-lying eigenvalues of the graph Laplacian for topologically trivial spacetimes are higher than they are for regular lattice causets, which is indeed what we observe. In addition, changes to the topology result in less connectivity, i.e., a tendency towards disconnected components, which significantly lowers the low-lying eigenvalues.

The only causet class that is more strongly connected at the lower end of the spectrum than topologically trivial spacetimes is the class of layered orders with a low number of layers, such as KR orders.

The highest-lying eigenvalue quantifies the layeredness of a causet. Therefore it equals two for layered  causets (meaning that they are exactly bipartite). It also allows to detect the layered structure in lattices, and the (pseudo-)layers in quasicrystals, by being significantly larger for those causets than for topologically trivial spacetimes.

\begin{figure}[!t]
    \centering
\includegraphics[width=\linewidth]{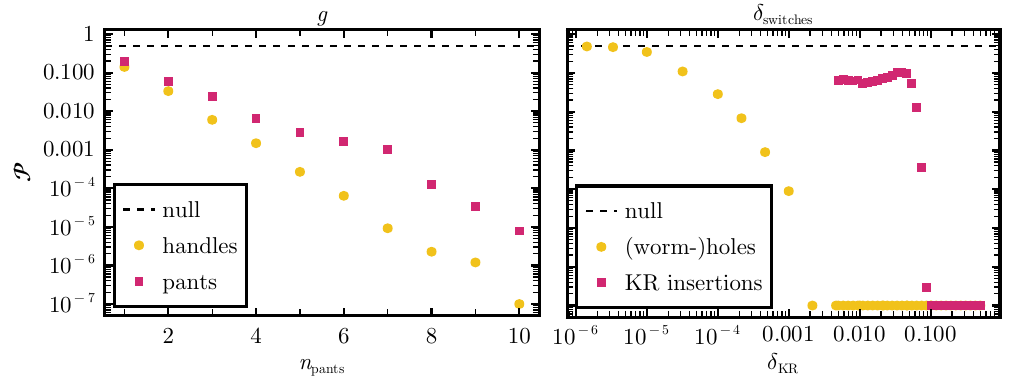}
\includegraphics[width=\linewidth]{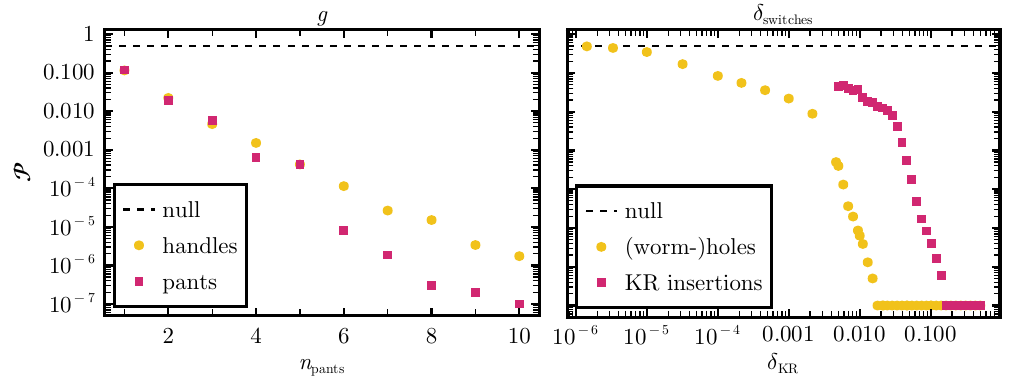}
    \caption{Distinguishability probabilities using the graph-Laplacian eigenvalues for spacetimes with handles, generalized-pants spacetimes, spacetimes with (worm-)holes and spacetimes with KR insertions, compared to the expected null value $1/2$ according to \cref{eq:distinguishability_probability} for ensembles of $10^3$ causets of size $\CSetSize=2048$ per data point ($10^4$ causets in the null class). The upper row is based on the full set of eigenvalues, the lower row only on the pair $(\LapEig_2,\LapEig_\CSetSize)$. The resolution for probabilities is $10^{-7}$, \ie, points with probability $10^{-7}$ are upper bounds rather than exact values.}
    \label{fig:evs_probabilities_panel}
\end{figure}

To compare quantitatively, we display distinguishability probabilities according to \cref{eq:distinguishability_probability} in \cref{tab:evs_distances} and \cref{fig:evs_probabilities_panel}. We report probabilities derived both from using the full set of $2048$ eigenvalues (for causets of size $\CSetSize=2048$) and from the strongly reduced set $(\LapEig_2,\LapEig_\CSetSize)$, which can be computed much more economically for large causets. Graph-Laplacian eigenvalues are by far the most effective at distinguishing different causet classes. Even the strongly reduced set $(\lambda_2,\lambda_\CSetSize)$ remains remarkably capable. For the full set of eigenvalues, all classes in \cref{tab:evs_distances} can be distinguished to a probability smaller than our datasets can resolve. Similarly, the eigenvalues detect modifications to the topology or boundary, or non-manifoldlikeness for far smaller modifications of the topologically trivial spacetimes than all other observables. This remains the case for the reduced set $(\LapEig_2,\LapEig_\CSetSize)$.

\begin{table}[!t]
    \centering
    \begin{tabular}{| l | c | c |}
    \hline\textbf{causet class} & \textbf{probability} & \textbf{probability (only $(\LapEig_2,\LapEig_\CSetSize)$)}\\\hline
    regular lattice causets & $<10^{-8}$ & $10^{-3}$\\\hline
    quasicrystal causets & $<10^{-8}$ & $<10^{-8}$\\\hline
    layered causets & $< 10^{-8}$ & $<10^{-8}$ \\\hline
    fixed-connectivity causets & $<10^{-8}$ & $10^{-5}$\\\hline
    \end{tabular}
    \caption{Distinguishability probabilities for the graph-Laplacian eigenvalues and their subset $(\LapEig_2,\LapEig_\CSetSize)$ for all causet classes according to \cref{eq:distinguishability_probability} for ensembles of $10^4$ elements of size $\CSetSize=2048$ per class.}
    \label{tab:evs_distances}
\end{table}

\subsection{Height profile}
\begin{figure}
    \centering
    \includegraphics[width=0.92\linewidth]{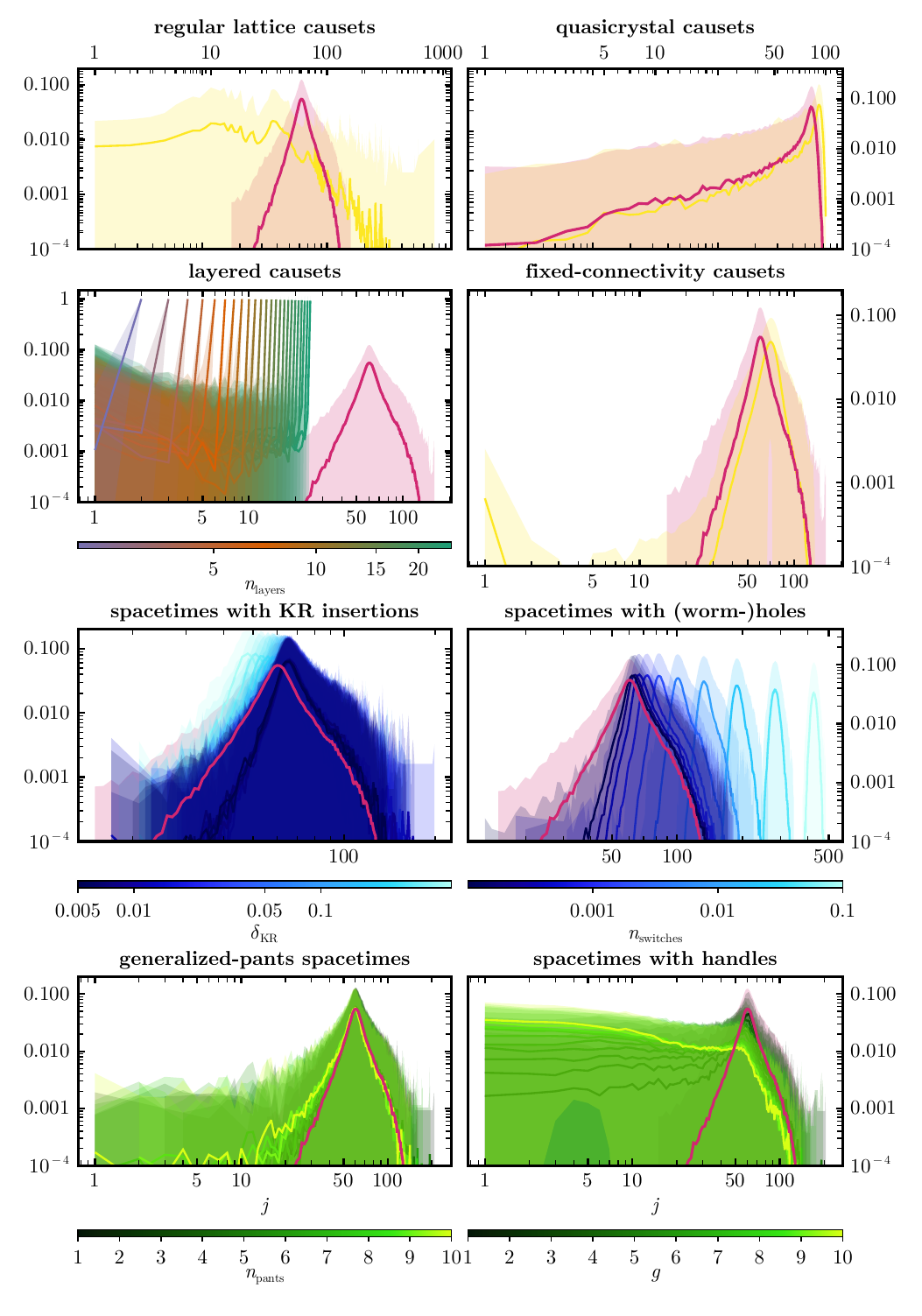}
    \caption{Height profile for different kinds of causal sets compared to sprinklings into topologically trivial spacetimes (magenta). Lines are means, bands are standard deviations. Datasets contain $10^4$ partial orders for each class ($3\cdot10^4$ for spacetimes with KR insertions and spacetimes with wormholes) of size 2048.}
    \label{fig:plot_matrix_height}
\end{figure}

Our final observable is the height profile, see \cref{fig:plot_matrix_height} for the results. It is mainly sensitive to the boundaries of a causal set (\cf \cref{app:bound_size_dim_height} for more details). In the task of reconstructing a continuum spacetime from a sprinkling, the height profile can be used to infer the shape of the boundary of the spacetime region into which the points have been sprinkled. 

We also note that the height profile is a rather noisy observable within most of the classes that we consider. Thus, small changes in the expectation value that arise, e.g., by increasing the number of pants or the ratio of KR insertions into a sprinkling, are too small to be robustly detectable.

Overall, Fig.~\ref{fig:plot_matrix_height} shows that, apart from constraining the shape of the boundary, the height profile is not a useful observable. Because the height profile is a very noisy observable, we refrain from making a more quantitative comparison.

\section{Conclusions and outlook\label{sec:conclusion}}
The configuration space of causal sets is vast and not well-understood beyond the observation that layered orders constitute (super-)exponentially dominant classes of causal sets \cite{Kleitman_1975, Henson:2016piq}. The number of causal sets with three layers is exponentially enhanced over those with four layers, which in turn is exponentially enhanced over those with five layers, and so on, at least when the causal set size is large. Little is known about the subdominant causal sets; their relative abundances as well as the observables that characterize and distinguish them. Observables inspired by continuum differential geometry and topology have been developed in \cite{vanderDuin:2025hmf,vanderDuin:2025ydn}, but have not been extensively tested for non-manifoldlike causal sets. In \cite{Surya:2025mvt}, the abundance of causal intervals was proposed as an observable that partially distinguishes sprinklings into Lorentzian manifolds.

In this paper, we make a step towards charting the configuration space by combining causal-set generation algorithms with the study of observables. Our aim is to achieve an understanding of the configuration space of causal sets in which causal sets are defined and simultaneously charted by means of the observables.

Observables inspired by continuum geometry, such as curvature invariants, are often very noisy \cite{Dowker:2013vba}, and expensive to calculate. In the case of the Ricci scalar curvature, it is even necessary to introduce an ad-hoc nonlocality scale which is many orders of magnitude larger than the discreteness scale to tame the fluctuations \cite{Sorkin:2007qi,Benincasa:2010ac,Dowker:2013vba}. Thus, continuum-geometry-inspired observables may not be the best candidates to distinguish manifoldlike from non-manifoldlike causal sets. Instead, we focus on label-invariant graph observables with no direct counterpart in continuum geometry. Label-invariance makes them ``good" observables in the sense that they do not break the discrete remnant of diffeomorphisms. However, they do not necessarily have physical meaning. Instead, they can be chosen specifically to have little noise, to be efficiently calculable and to distinguish between different classes of causal sets.

To quantify the distinguishability of classes of causal sets by a given observable, we follow a probabilistic approach: We calculate the intrinsic variability of an observable across a single comparison class of causal sets. We compare this to the typical difference between the observable evaluated in a different class and the comparison class. If the difference is larger than the intrinsic variability, the observable can distinguish between the two classes.

The four observables that we have identified (link degree distribution, interval abundance, eigenvalues of the graph Laplacian and height profile) are, in combination, capable of distinguishing the classes of causal sets we have considered here, at least for the size, dimensionality and boundary type that we have focused on.
The height profile is mainly sensitive to global properties and is a noisy observable. Thus, it can probably be dropped from the set of observables. The other three observables show little noise for the topologically trivial sprinklings. In particular, the graph-Laplacian eigenvalues, even the strongly reduced set $(\LapEig_2,\LapEig_\CSetSize)$ alone, can distinguish well between all the classes we considered. This enables us to distinguish between the nine different classes of causal sets that we study here.

We have also explored the recently introduced Lorentzian quasicrystals \cite{Boyle:2026vgo}. Because they do not select a preferred frame, one may have considered them as a viable alternative to sprinklings. We do, however, find that the interval abundance is sensitive to the underlying higher-dimensional periodic structure which breaks Lorentz invariance in the higher-dimensional space. If one requires that no Lorentz-breaking structures, not even auxiliary ones such as the four-dimensional lattice, should leave imprints on the observables, our results disfavor quasicrystals as physically relevant discrete spacetimes. Nevertheless, they remain a part of the general causal-set configuration space, just like regular lattices. They can, however, be relatively straightforwardly distinguished from sprinklings.

For future work, there are several pertinent questions, which we list in roughly increasing level of difficulty.\\
First, we have not investigated the effects of varying the causal-set size $n$ on the distinguishability of different classes of causal sets. As larger causets resolve geometric and topological features better, we expect the classes to be even more distinguishable at large causal-set size.\\
Second, the effects of choosing different boundaries for those causets that can be embedded, and how they impact the distinguishability of different classes can be investigated in more detail than we did here.\\
Third, we have grouped causets into classes according to the algorithm by which they are generated. However, for some of the classes, it is also of interest to understand whether observables can distinguish sub-classes within a given class. For instance, for the topologically trivial spacetimes, we are sampling over different, and locally varying, conformal factors. Thus, this class contains sprinklings into manifolds with very different average values of curvature and amount of variation of curvature, connected to the presence or absence of Killing vectors. Exploring  within this class along the lines suggested in \cite{Surya:2025mvt}, where the interval abundance serves to define a distance on the space of Lorentzian geometries, and understanding which different conformally flat two-dimensional geometries have equal interval abundance, is clearly of interest. From the relatively small standard deviation that our result for the interval abundance exhibits, we might expect that these geometries have (nearly) equal interval abundances -- a question that clearly deserves more investigation. \\
Fourth, sprinkling into higher spacetime dimensions is clearly of relevance, although computationally significantly more expensive. In higher spacetime dimensions, we also have access to spacetimes with other nontrivial curvature invariants besides the Ricci scalar and its derivative. So far, in causal sets only expressions for discrete counterparts of $R$ \cite{Benincasa:2010ac} and $\Box R$ \cite{deBrito:2023axj} are known. Therefore, understanding which graph observables are sensitive to changes in, e.g., the Kretschmann scalar, may provide a hint towards the construction of a discrete counterpart of this curvature invariant.\\
Fifth, the set of classes of causets can be enlarged. For instance, we have not included causets generated by a dynamical growth process \cite{Rideout:1999ub,Rideout:2000fh,Varadarajan:2005gg,Dowker:2017zqj,Dowker:2019qiz,Surya:2020cfm,Zalel:2020oyf,Bento:2021voo,Bento:2021mev,Zalel:2023uwy}, which clearly constitute another relevant class for future studies.\\
Sixth, one can refine our probabilistic measure of distinguishability to quantify whether a single causal set belongs to a certain reference class instead of comparing two classes, \eg, whether a single causal set is manifoldlike. Within-class variation in the reference class sets the scale for any pairwise histogram distance. Then, one can express the histogram distance of an observable between the single causal set and the average of the reference class in units of this scale. Accordingly, one can quantify how probable this single causet is to be a member of the reference class, when considering the information captured by the observable. The power of this ansatz is apparent when considering several observables independently. This allows not only to capture that, say, a causet is non-manifoldlike, but also to characterize in which way it deviates from a typical sprinkling.\\
Seventh, the problem of distinguishing different classes of causal sets appears tailor-made for machine learning algorithms. Once a comprehensive set of complementary, sensitive and easy-to-compute observables is compiled, these algorithms should refine the observables even further. As a result, characterizing a causal set, including its geometric and topological properties for manifoldlike causal sets, may be within reach. As a first step, the different classes of causal sets that we have defined can be used as a training set for machine-learning algorithms that learn to distinguish topologically trivial spacetimes from other causal sets, in particular from non-manifoldlike ones.\\
Eighth, we have defined two classes of causal sets for which we have formulated the expectation that they coarse-grain to manifoldlike causal sets, namely sprinklings with KR insertions and spacetimes with (worm-)holes. There is, however, no well-established coarse-graining process for causal sets. Standard ``block-spinning" type of transformations are not straightforward to define, because they rely on spatially localized regions, whereas nearest neighbors in spacetime do not form local regions. Developing a suitable Lorentzian coarse-graining procedure is clearly a question of much broader relevance in quantum gravity.

\appendix

\crefalias{section}{appsec}

\section{Link-degree distribution of topologically trivial spacetimes \label{app:LinkDistMan}}

In this appendix, we examine the link-degree distribution of topologically trivial spacetimes more precisely. We characterize the distribution with its first moments and use them to substantiate that the link-degrees of topologically trivial spacetimes are Gaussian distributed up to pure boundary effects. 

We define the expectation value of random variable $\RandomVariable(j)$ with respect to the link-degree distribution $\DegreeDist{j}(C)$ as 
\begin{equation}
    \ExpectationValue{X(\HistDummy)}_{\DegreeDist{j}(C)} \equiv \sum_j X(j)\,\DegreeDist{j}(C).
\end{equation}
We focus on the mean $\Mean(C)$, standard deviation $\Std(C)$, skewness $\Skewness(C)$ and excess kurtosis $\Kurtosis(C)$,
\begin{eqnarray}
\Mean(C)&\equiv&\ExpectationValue{j}_{\DegreeDist{j}(C)},\\
\Std(C)&\equiv&\ExpectationValue{(j-\mu)^2}_{\DegreeDist{j}(C)},\\
\Skewness(C)&\equiv&\frac{\ExpectationValue{(j-\mu)^3}_{\DegreeDist{j}(C)}}{\Std^{3/2}},\\
\Kurtosis(C)&\equiv&\frac{\ExpectationValue{(j-\mu)^4}_{\DegreeDist{j}(C)}}{\Std^{2}}-3.
\end{eqnarray}
We further define the ensemble mean and standard deviation over skewness and excess kurtosis as
\begin{align}
    \gamma_i=\frac{1}{|\Ensemble|}\sum_{\CSet\in \Ensemble}\gamma_i(\CSet), &&\Std_{\Ensemble}(\gamma_i)=\frac{1}{|\Ensemble|}\sum_{\CSet\in \Ensemble}\left(\gamma_i(\CSet)- \gamma_i\right)^2,
\end{align}
for $i=1,2$. A Gaussian distribution has $\Skewness=\Kurtosis=0$. For topologically trivial spacetimes of size $\CSetSize=2048$, the skewness and excess kurtosis are $(\gamma_1,\Std_\Ensemble(\Skewness))=(0.24,0.08)$ and $(\gamma_2,\Std_\Ensemble(\Kurtosis))=(0.0,0.2)$, respectively. Thus, while the degree distribution is skewed towards smaller degrees, the excess kurtosis indicates that the shape of its peak is Gaussian.

\begin{figure}[!t]
    \centering
    \begin{minipage}{.49\linewidth}
        \includegraphics[width=\linewidth]{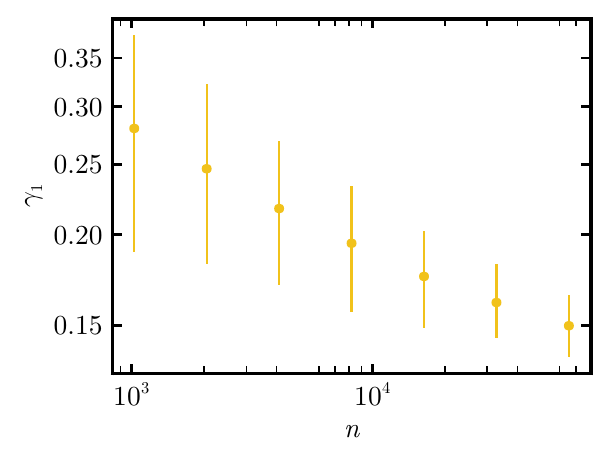}
    \end{minipage}
    \begin{minipage}{.49\linewidth}
        \includegraphics[width=\linewidth]{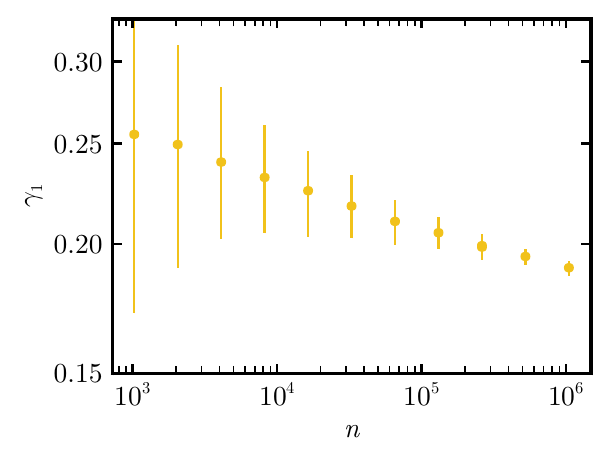}
    \end{minipage}
    \caption{Skewness $\Skewness$ of the link-degree distribution as a function of causet size $\CSetSize$ for topologically trivial spacetimes with rectangular boundary (left) and Minkowski sprinklings with causal-diamond boundary (right), averaged over ensembles of $10^3$ causets per data point ($10^2$ causets per data point for $\CSetSize\geq5\cdot10^5$. The asymmetric error bars reflect the 16th and the 84th percentile (for Gaussian distributions: $\pm1\Std$).}
\label{fig:degree_skew_size_scaling}
\end{figure}

We now give evidence that the skewness is a pure boundary effect. We show $\Skewness$ as a function of causet size $\CSetSize$ in \cref{fig:degree_skew_size_scaling} for topologically trivial spacetimes with rectangular boundary and Minkowski sprinklings with causal-diamond boundary. For both boundary types, the ensemble mean $\Skewness$ decays as a function of $\CSetSize$, and errors decrease. Thus, the size dependence is compatible with $\lim_{\CSetSize\to\infty}\Skewness(\CSetSize)\to0$, even though we cannot exclude a finite limit. 

Further support comes from Minkowski quasicrystals with causal-diamond boundary, whose link-degree distribution has skewness $(\Skewness,\Std_\Ensemble(\Skewness))_{\rm cryst}=(0.27,0.05)$. This is compatible with the skewness for Minkowski sprinklings with causal-diamond boundary $(\Skewness,\Std_\Ensemble(\Skewness))_{\rm Mink}=(0.25,0.06)$. Thus, Lorentz invariant causets of equal boundary have an equally skewed link-degree distribution. Quasicrystal causets, in turn, do not have Gaussian distributed link degrees. Their link-degree distribution is significantly more peaked than a Gaussian such that $(\Kurtosis,\Std_\Ensemble(\Kurtosis))_{\rm cryst}=(0.6,0.1)$. 

\section{Size, boundary and dimensionality dependence of the observables\label{app:bound_size_dim}}

In this appendix, we summarize how the link-degree distribution, the interval abundances, the graph-Laplacian eigenvalues and the height profile change when varying causet size, boundary or dimensionality. We only consider topologically trivial spacetimes, but varying curvature.

\subsection{Link-degree distribution}

\begin{figure}[!t]
    \centering
    \includegraphics[width=\linewidth]{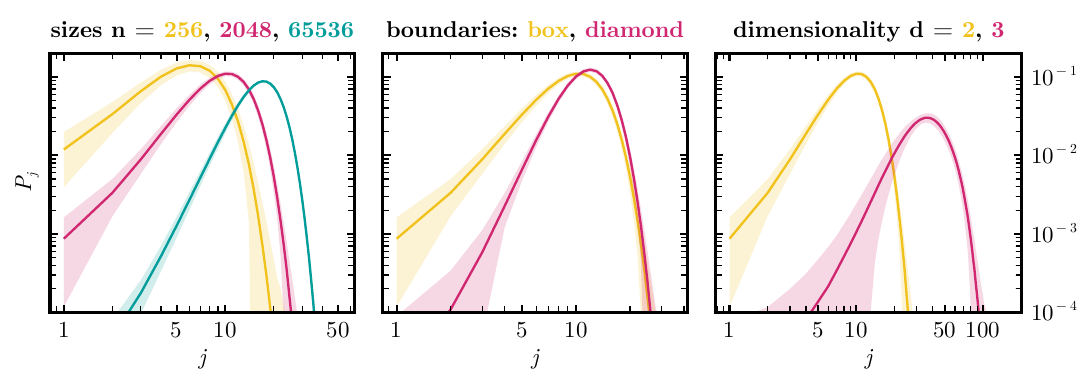}
    \caption{Degree distribution for manifoldlike causets of trivial topology with varying causal-set size (left panel), boundary type (central panel) and spacetime dimensionality (right panel). Each dataset contains $10^4$ causets of size $\CSetSize=2048$ with rectangular boundary and $\Dim=2$ unless specified otherwise. We plot the means as lines and the standard deviations as bands.
   }
\label{fig:degree_size_boundary_dimension}
\end{figure}

We show how the link-degree distribution varies in \cref{fig:degree_size_boundary_dimension}. All three changes shift the maximum of the link distribution to larger values. Further, the distribution becomes less skewed. A change in the boundary type from a box-boundary to a causal-diamond boundary changes the relative number of elements close
to the boundary and thus reduces the relative abundance of elements with few links. A first comparison to Fig.~\ref{fig:plot_matrix_degrees} shows that changes in causal-set type affect the link distribution in dissimilar ways.

\subsection{Interval abundances}

\begin{figure}
    \centering
    \includegraphics[width=\linewidth]{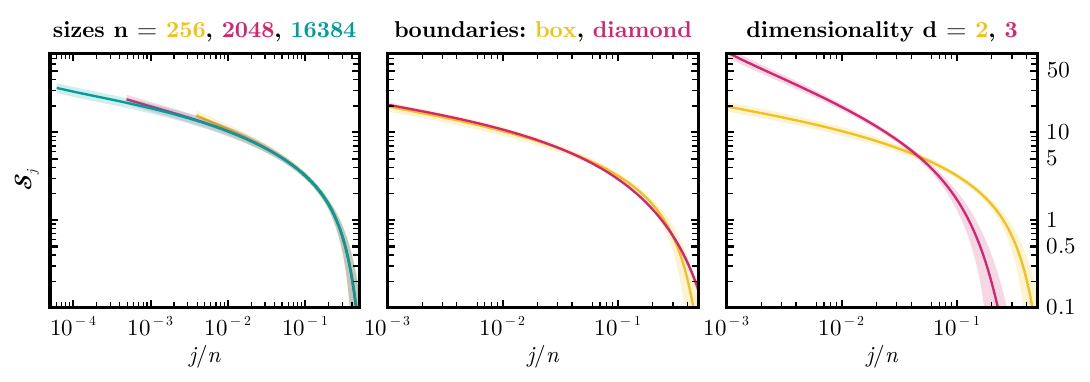}
    \caption{Interval abundances for manifoldlike causets of trivial topology with varying causal-set size (left panel), boundary type (central panel) and spacetime dimensionality (right panel). Each dataset contains $10^4$ causets of size $\CSetSize=2048$ with rectangular boundary and $\Dim=2$ unless specified otherwise. We plot the means as lines and the standard deviations as bands.}
    \label{fig:cardinalities_size_boundary_dimension}
\end{figure}

We examine how the interval abundances vary in \cref{fig:cardinalities_size_boundary_dimension}. We find that with our normalization, the interval abundances are largely size- and boundary-independent. The dimensionality, however, has a very strong imprint leading to many more small intervals and fewer large intervals. This effect can be clearly differentiated from the features of the other causal-set classes we study.

\subsection{Graph-Laplacian eigenvalues\label{app:bound_size_dim_lap_eig}}

\begin{figure}
    \centering
    \includegraphics[width=\linewidth]{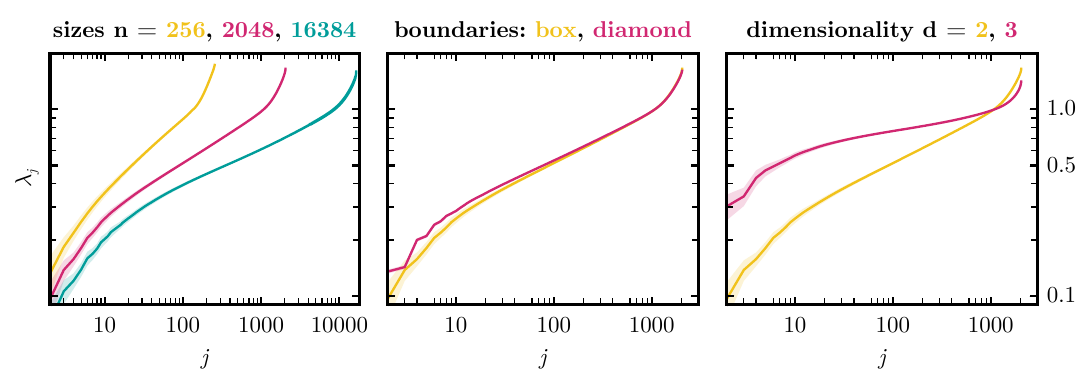}
    \caption{Graph-Laplacian eigenvalues for manifoldlike causets of trivial topology with varying size, boundary and dimensionality. Every single dataset contains $10^4$ causets of size $\CSetSize=2048$ with rectangular boundary and $\Dim=2$ unless specified otherwise. We plot the means as lines and the standard deviations as bands. Left: varying size. Middle: Manifoldlike causets with trivial topology in rectangular boundary compared to Minkowski sprinklings in causal-diamond boundary. Right: varying dimensionality.}
    \label{fig:ev_sym_link_size_boundary_dimension}
\end{figure}

We summarize how the graph-Laplacian eigenvalues vary in \cref{fig:ev_sym_link_size_boundary_dimension}. The eigenvalues generally decrease with increasing causet size. This indicates that large causets have larger, more loosely connected subregions, and are less layered. While this reflects the natural inverse square scaling of the eigenvalues of geometric Laplacians with region size, it is clearly not the case for all causet types. The largest eigenvalue of layered orders always equals 2, independent of causet size. Low-lying eigenvalues, which are sensitive to large-scale connectivity, are boundary dependent, while the larger ones are not. Finally, the Laplacian eigenvalues are very sensitive to the dimensionality. Higher-dimensional topologically trivial spacetimes are significantly more connected at intermediate scales (large low-lying eigenvalues), while being less connected at small scales and less bipartite (small high-lying eigenvalues). These changes can be clearly distinguished from those introduced by considering different causet types.

\subsection{Height profile\label{app:bound_size_dim_height}}

\begin{figure}
    \centering
    \includegraphics[width=\linewidth]{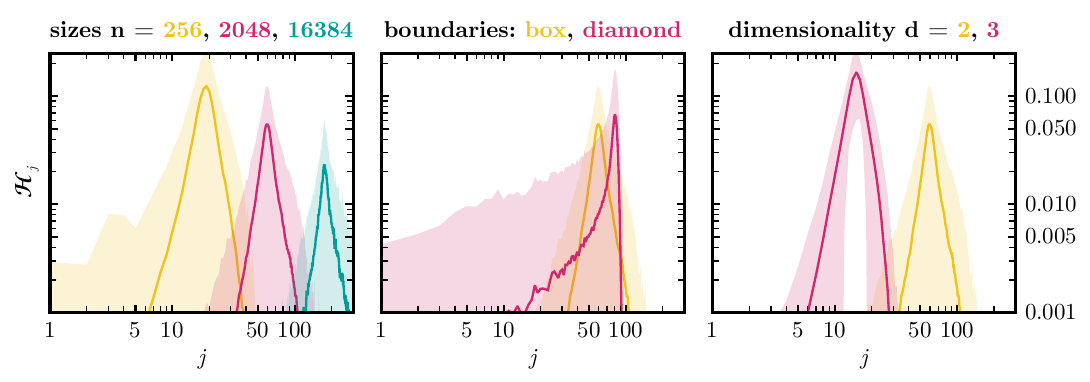}
    \caption{Height profile for manifoldlike causets of trivial topology with varying parameters. Every single dataset contains $10^4$ causets of size $\CSetSize=2048$ with rectangular boundary and $\Dim=2$ unless specified otherwise. We plot the means as lines and the standard deviations as bands. Left: varying size. Middle: Manifoldlike causets with trivial topology in rectangular boundary compared to Minkowski sprinklings in causal-diamond boundary. Right: varying dimensionality.}
    \label{fig:height_size_boundary_dimension}
\end{figure}

 We show how the height profile varies in \cref{fig:height_size_boundary_dimension}. The causal-diamond height profile clearly has a different shape compared to a rectangular boundary, showing how sensitive the height profile is to boundary modifications. Besides, the height profile is shifted towards larger heights for larger linear extent in one dimension $\CSetSize^{1/\Dim}$, and becomes less noisy with increasing dimensionality. 

\section{Convergence plots\label{app:convergence_plots}}

In this appendix, we discuss in which way the observables computed with our datasets converge to the values we show in \cref{fig:plot_matrix_degrees,fig:plot_matrix_cardinalities,fig:plot_matrix_eigvals,fig:plot_matrix_height}. We discuss the link-degree distribution, the interval abundances, the graph-Laplacian eigenvalues and the height profile for topologically trivial spacetimes. 

Our strategy is as follows: We randomly reduce the datasets to some sample size $\SampleSize$, and compute the mean and the standard deviation of observables as we successively increase $\SampleSize$ to the dataset size $\SampleSize_{\rm max}$. For each bin $\HistDummy$ of our observable $\Observable_j$, we define the mean $\Mean_{\Observable_j}(\SampleSize)$ and the standard deviation $\Std_{\Observable_j}(\SampleSize)$. We fit power laws to the distance of $\Mean_{\Observable_j}(\SampleSize)$ and $\Std_{\Observable_j}(\SampleSize)$ from their (approximated) values at infinite sample size 
\begin{equation}
    \mu_{\Observable_j,\infty}\equiv\lim_{\SampleSize\to\infty}\Mean_{\Observable_j}(\SampleSize)\simeq\Mean_{\Observable_j}(\SampleSize_{\rm max}),\qquad \Std_{\Observable_j,\infty}\equiv\lim_{\SampleSize\to\infty}\Std_{\Observable_j}(\SampleSize)\simeq\Std_{\Observable_j}(\SampleSize_{\rm max}).
\end{equation}
By showing that the best-fit power law decays, we make sure that we are in the convergent regime. The fit evaluated at $\SampleSize_{\rm max}$ then provides the error to the mean and the standard deviation of the observable. Topologically trivial spacetimes intrinsically vary with the parameters $\ChebyCoeff_{ij}$ and $\ChebyDecayBase$ we provided in \cref{tab:manifoldlike_params}, \ie, with the curvature of the geometry we sprinkle into. Therefore, generally $\Std_{{\Observable_j},\infty}\neq 0$.

More precisely, with successively increasing sample size $\SampleSize$, we fit
\begin{align}
    |\Mean_{\Observable_j}(\SampleSize)-\Mean_{\Observable_j,\infty}|=&\MeanFitFac_{\Observable_j} \SampleSize^{-\MeanFitExp_{\Observable_j}},\\
    |\Std_{\Observable_j}(\SampleSize)-\Std_{\Observable_j,\infty}|=&\StdFitFac_{\Observable_j} \SampleSize^{-\StdFitExp_{\Observable_j}},
\end{align}
where $\StdFitFac_{\Observable_j},$ $\MeanFitFac_{\Observable_j},$ $\StdFitExp_{\Observable_j}$ and $\MeanFitExp_{\Observable_j}$ are fit parameters. In a least squares fit approximating
\begin{align}
    \Min_{\MeanFitFac_{\Observable_j}, \MeanFitExp_{\Observable_j}}\sum_{\SampleSize}\left(|\Mean_{\Observable_j}(\SampleSize)-\Mean_{\Observable_j,\infty}|-\MeanFitFac_{\Observable_j}\SampleSize^{-\MeanFitExp_{\Observable_j}}\right)^2,
\end{align}
we can analytically minimize with respect to $\MeanFitFac_{\Observable_j}$, obtaining
\begin{equation}
    \MeanFitFac_{\Observable_j}=\frac{\sum_\SampleSize|\Mean_{\Observable_j}(\SampleSize)-\Mean_{\Observable_j,\infty}|\SampleSize^{-\MeanFitExp_{\Observable_j}}}{\sum_\SampleSize N^{-2\MeanFitExp_{\Observable_j}}}.
\end{equation}
Therefore, we only have to fit the decay exponents $\MeanFitExp_{\Observable_j}$. As a result, we numerically optimize
\begin{align}
    \Min_{\MeanFitExp_{\Observable_j}}\sum_\SampleSize\left(|\Mean_{\Observable_j}(\SampleSize)-\Mean_{\Observable_j,\infty}|-\frac{\sum_{\SampleSize'}|\Mean_{\Observable_j}(\SampleSize')-\Mean_{\Observable_j,\infty}|(\SampleSize')^{-\MeanFitExp_{\Observable_j}}}{\sum_{\SampleSize''} (N'')^{-2\MeanFitExp_{\Observable_j}}}\SampleSize^{-\MeanFitExp_{\Observable_j}}\right)^2,
\end{align}
to find the decay exponents, and with them the errors of the means of each bin of $\Observable$. Following the same derivation, we optimize
\begin{align}
    \Min_{\StdFitExp_{\Observable_j}}\sum_\SampleSize\left(|\Std_{\Observable_j}(\SampleSize)-\Std_{\Observable_j,\infty}|-\frac{\sum_{\SampleSize'}|\Std_{\Observable_j}(\SampleSize')-\Std_{\Observable_j,\infty}|(\SampleSize')^{-\StdFitExp_{\Observable_j}}}{\sum_{\SampleSize''} (\SampleSize'')^{-2\StdFitExp_{\Observable_j}}}\SampleSize^{-\StdFitExp_{\Observable_j}}\right)^2,
\end{align}
to obtain the best-fit decay coefficient and error for the standard deviations.

Given a fit, we can estimate the relative amount to which the mean of the observable has converged, \ie, the relative convergence error as
\begin{equation}
    \delta\mu_{\Observable_j} = \frac{\MeanFitFac_{\Observable_j}\SampleSize_{\rm max}^{-\MeanFitExp_{\Observable_j}}}{\Mean_{\Observable_j}(\SampleSize_{\rm max})}.\label{eq:rel_conv_error}
\end{equation}
To estimate the goodness of the fit, we define the residual at each $\SampleSize$
\begin{equation}
    {\rm Res}_{\Mean_{\Observable_j}}(\SampleSize) \equiv |\Mean_{\Observable_j}(\SampleSize)-\Mean_{\Observable_j,\infty}|-\MeanFitFac_{\Observable_j}|\SampleSize^{-\MeanFitExp_{\Observable_j}}.
\end{equation}
Then, we can define the normalized root mean square
\begin{equation}
    {\rm NRMS}_{\Mean_{\Observable_j}}\equiv\frac{\sqrt{\frac{1}{\SampleSize_{\rm max}}\sum_{\SampleSize=1}^{\SampleSize_{\rm max}}\left({\rm Res^{\Observable_j}_\Mean(\SampleSize)}\right)^2}}{\Mean_{\Observable_j}(\SampleSize_{\rm max})}.\label{eq:NRMS}
\end{equation}
This quantifies the error that the fit makes per point in comparison to the data relative to the final value of the mean (or standard deviation) of the observable. We define a fit to the randomized data to be acceptable if ${\rm NRMS}_{\Mean_{\Observable_j}}\leq10\%$.

\begin{figure}
    \centering
    \includegraphics[width=.895\linewidth]{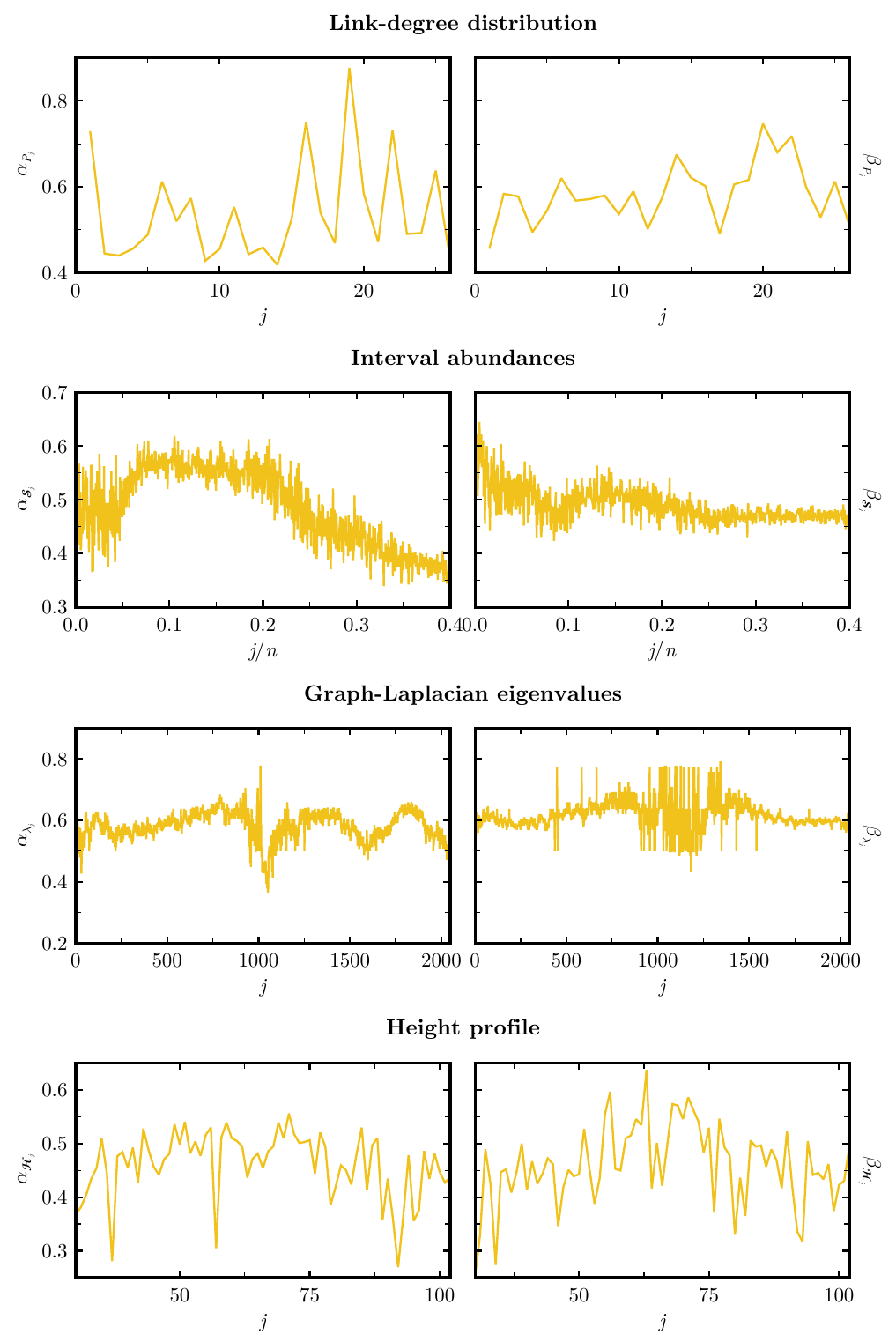} 
    \caption{Scaling exponents for sample-size scaling of means and standard deviations of the considered graph observables for topologically trivial spacetimes of size $2048$.}
    \label{fig:convergence_plot_matrix}
\end{figure}

We show the resulting scaling exponents for all considered bins of all considered observables in \cref{fig:convergence_plot_matrix}. They are all positive, implying that both means and standard deviations are convergent. In addition, we show the minimal-bin and maximal-bin $\rm NRMS$ and relative convergence errors for means and standard deviations of all graph observables in \cref{tab:convergence}. As expected, intrinsically less noisy observables like the graph-Laplacian eigenvalues have converged to a far greater extent than, for example, the height profile. Nevertheless, the observables have converged sufficiently in all bins shown in \cref{fig:convergence_plot_matrix}, which we used for the plots of topologically trivial spacetimes in \cref{fig:plot_matrix_degrees,fig:plot_matrix_cardinalities,fig:plot_matrix_eigvals,fig:plot_matrix_height}.

\begin{table}
    \centering
    \begin{tabular}{@{}r@{\quad}|c|c|c|c||c|c|c|c|}
    \cline{2-9}
        &  \multicolumn{2}{c|}{${\rm NRMS}_{\Mean_{\Observable_j}}$}
        & \multicolumn{2}{c||}{$\delta\Mean_{\Observable_j}$}
        & \multicolumn{2}{c|}{${\rm NRMS}_{\Std_{\Observable_j}}$}
        & \multicolumn{2}{c|}{$\delta\Std_{\Observable_j}$} \\
        \cline{2-9}
        & $\min_\HistDummy$ & $\max_\HistDummy$
        & $\min_\HistDummy$ & $\max_\HistDummy$
        & $\min_\HistDummy$ & $\max_\HistDummy$
        & $\min_\HistDummy$ & $\max_\HistDummy$ \\
        \cline{2-9}
        Link-degree distribution      & $0.06\%$ & $5\%$ & $0.02\%$ & $4\%$ & $0.7\%$ & $3\%$ & $0.1\%$ & $3\%$ \\
        Interval abundances           & $0.01\%$ & $1\%$  & $0.01\%$ & $1\%$ &$1\%$ & $2\%$ & $0.3\%$ & $2\%$\\
        Graph-Laplacian eigenvalues   &  $0.04\%$ & $0.4\%$ & $0.01\%$ & $0.2\%$ & $0.9\%$ & $2\%$ & $0.09\%$ & $0.2\%$ \\
        Height profile           & $1\%$ & $10\%$  & $0.6\%$ & $1\%$ & $1\%$ & $9\%$ & $0.3\%$ & $11\%$ \\
        \cline{2-9}
    \end{tabular}
    \caption{Minimal-bin and maximal-bin normalized root mean square (\cf \cref{eq:NRMS}) and relative convergence error (\cf \cref{eq:rel_conv_error}) of the means and standard deviations of the considered observables for ensembles of $10^4$ topologically trivial spacetimes of size $2048$. Only the bins plotted in \cref{fig:convergence_plot_matrix} have been considered.}
    \label{tab:convergence}
\end{table}

\FloatBarrier

\bibliographystyle{utphys}
\bibliography{refs}

\end{document}